\documentclass[a4paper,11pt]{article}
\usepackage{jheppub} % for details on the use of the package, please see the JINST-author-manual
\usepackage{lineno}
\usepackage[vcentermath]{youngtab}
\usepackage{graphicx} 
\usepackage{url}
\usepackage{hyperref}
\usepackage{amsmath}
\usepackage{amssymb}
\usepackage{outlines}
\usepackage{color}
\usepackage{hyperref}
\usepackage{tikz}
\usetikzlibrary{trees}
\usepackage{comment}

\newcommand{\cO}{\mathcal O}

\newcommand{\be}{\begin{equation}}
\newcommand{\ee}{\end{equation}}
\newcommand{\bea}{\begin{eqnarray}}
\newcommand{\eea}{\end{eqnarray}}
\newcommand{\ba}{\begin{equation} \begin{aligned}}
\newcommand{\ea}{\end{aligned} \end{equation}}

\newcommand{\Df}{\Delta_\phi}

\newcommand{\De}{\Delta}

\newcommand{\approptoinn}[2]{\mathrel{\vcenter{
  \offinterlineskip\halign{\hfil$##$\cr    #1\propto\cr\noalign{\kern2pt}#1\sim\cr\noalign{\kern-2pt}}}}}

\makeatletter
\def\@fpheader{\relax}
\makeatother

\title{\boldmath Moments and saddles of heavy CFT correlators}
\author{David Poland, Gordon Rogelberg}
\affiliation{Department of Physics, Yale University, 217 Prospect St, New Haven, CT 06520, USA}
\emailAdd{david.poland@yale.edu, gordon.rogelberg@yale.edu}

\abstract{
We study the operator product expansion (OPE) of identical scalars in a conformal four-point correlator as a Stieltjes moment problem, and use Riemann-Liouville type fractional differential operators to generate classical moments from the correlation function. We use crossing symmetry to derive leading and subleading relations between moments in $\Delta$ and $J_2 \equiv \ell(\ell+d-2)$ in the ``heavy" limit of large external scaling dimension, and combine them with constraints from unitarity to derive two-sided bounds on moment sequences in $\Delta$ and the covariance between $\Delta$ and $J_2$. The moment sequences which saturate these bounds produce ``saddle point" solutions to the crossing equations which we identify as particular limits of correlators in a generalized free field (GFF) theory. This motivates us to study perturbations of heavy GFF four-point correlators by way of saddle point analysis, and we show that saddles in the OPE arise from contributions of fixed-length operator families encoded by a decomposition into higher-spin conformal blocks. To apply our techniques, we consider holographic correlators of four identical single scalar fields perturbed by a bulk interaction, and use their first few moments to derive Gaussian weight-interpolating functions that predict the OPE coefficients of interacting double-twist operators in the heavy limit.}

\begin{document}
\maketitle
\pagenumbering{roman}
\setcounter{page}{2}

\newpage
\pagenumbering{arabic}
\setcounter{page}{1}
\section{Introduction}
\label{sec:intro}

The conformal bootstrap~\cite{Ferrara:1973yt, Polyakov:1974gs,Poland:2018epd} aims to constrain or even solve conformal field theories (CFT) by systematically imposing consistency conditions and symmetries. CFTs not only describe universality classes of systems at their second-order phase transitions, but they also describe the space of asymptotic observables for a quantum field theory (QFT) in Anti de Sitter (AdS) space of one dimension higher~\cite{Maldacena:1997re,Gubser:1998bc,Witten:1998qj}. In the AdS/CFT correspondence, the conserved stress tensor on the boundary CFT is dual to a bulk graviton, allowing us to study theories of quantum gravity by probing their dual CFT. 

The structure of a CFT arises from its convergent and associative operator product expansion (OPE). By performing appropriate conformal transformations, we can bring two local operators arbitrarily close to each other so that their product can be decomposed into an infinite number of primary operators of the form
\ba
\cO_i(0)\cO_j(x) = \sum_{\cO_k} \frac{\lambda_{\cO_i \cO_j \cO_k}}{\lambda_{\cO_k\cO_k}}C_{ijk}(x,\partial_x) \cO_k(x),
\ea
where $\lambda_{\cO_i \cO_j \cO_k}$ are OPE coefficients extracted from the normalization of a three-point correlator, $\lambda_{\cO_k\cO_k}$ is the normalization of two point correlators $\langle \cO(0) \cO(x) \rangle = \lambda_{\cO\cO} x^{-2\Delta_\cO}$, and $C_{ijk}(x,\partial_x)$ is a differential operator satisfying

\ba
\langle \cO_i(0)\cO_j(x)\cO_k(y)\rangle = \lambda_{\cO_i \cO_j \cO_k}C_{ijk}(x,\partial_x) (x - y)^{-2\Delta_k}.
\ea

Considering a conformal four-point function of identical scalars, we can take the OPE between two pairs of operators and decompose it as
\ba
\langle \phi(0) \phi(z,\bar{z}) \phi(1) \phi(\infty) \rangle  = \sum_\cO \frac{\lambda^2_{\phi\phi\cO}}{\lambda^2_{\cO\cO}} C_{\phi\phi\cO}(z,\partial_z)C_{\phi\phi\cO}(\bar{z},\partial_{\bar{z}} ) 
 \langle \cO(0) \cO(z,\bar{z}) \rangle.
\label{4ptope}
\ea
We identify $C_{\phi\phi\cO}(z,\partial_z)C_{\phi\phi\cO}(\bar{z},\partial_{\bar{z}})\langle \cO(0) \cO(z,\bar{z}) \rangle = \lambda_{\cO\cO}G_{\Delta,\ell}(z,\bar{z})$ as a conformal block, parametrized by the scaling dimension $\Delta$ and spin $\ell$ quantum numbers of $\cO$. In Euclidean signature $\bar{z} = z^*$, while in Lorentzian signature $z$ and $\bar{z}$ are independent real numbers. In general, the analytic continuation of the block maps $(z,\bar{z}) \in \mathcal{R}^2 \to \mathbb{C} $, where $\mathcal{R} = \mathbb{C}/( (-\infty,0]\bigcup [1,\infty))$ is the double cut plane. The conformal block is a group harmonic which resums the contributions of an irreducible representation, labeled by its lowest-weight (or ``primary") vector $\cO$, to the correlation function. The ability to produce such a decomposition is a consequence of Plancherel's theorem for the conformal group~\cite{bams/1183531812}. Most importantly, this decomposition allows us to describe any four-point correlator by a countable set of ``CFT data," which consists of the spectrum $\{\cO\}$ and OPE coefficients $\{\lambda_{\phi\phi\cO} \}$. Taking a union of these data for all four-point correlators in a given theory then uniquely describes all the local observables of the CFT.\footnote{Holographically, the OPE encodes the distribution of intermediate states exchanged in a scattering process through the AdS bulk, where each term in the sum of eq.~(\ref{4ptope}) may be replaced by a geodesic Witten exchange diagram~\cite{Hijano:2015zsa}.}

An important constraint arises from the associativity of the OPE, where we can equate decompositions of the correlator in different channels, corresponding to different choices of pairs of operators. This property gives rise to the s-t channel ``crossing equation" 
\ba
\sum_\cO a_\cO F_{\Delta,\ell}(u,v) = 0,
\ea
where $a_\cO =\frac{ \lambda^2_{\phi\phi\cO}}{\lambda_{\phi\phi}^2 \lambda_{\cO\cO}}$ is the squared and normalized OPE coefficient, 
\ba
F_\cO(u,v)  = u^{-\Df} G_{\Delta,\ell}(u,v) - v^{-\Df} G_{\Delta,\ell}(v,u)
\ea
is the crossing vector, and $u=z \bar{z}, \, v= (1-z)(1-\bar{z})$ are the standard conformal cross ratios. Another important constraint on the decomposition arises for unitary CFTs, and imposes that the OPE coefficients are real so that all $a_\cO$ are positive. Applying a basis of functionals to this sum rule and using the positivity of $a_\cO$ allows one to rule out certain CFT spectra by preparing a functional that produces a contradiction after acting on the proposed spectrum. Functionals which are constructed to prove an optimal bound such as the maximum allowed scalar gap or a given OPE coefficient are called ``extremal functionals," and encode the spectrum of the correlator which saturates such a bound in their root structure. One can implement the search for such a functional as a semi-definite program (SDP) which can be solved numerically~\cite{El-Showk:2012vjm}. 

In the past decade, there has been tremendous growth in the numerical conformal bootstrap program yielding crucial insights into the structure of CFTs. Notably, we can now compute precise quantum numbers of a large number of operators in the 3d Ising CFT~\cite{El-Showk:2012cjh,El-Showk2014-ce,Kos:2016ysd,Chang:2024whx}, the O(N) vector models~\cite{Kos:2013tga,Chester:2019ifh,Chester:2020iyt}, Gross-Neveu-Yukawa CFTs~\cite{Atanasov:2022bpi, Erramilli2022-yi}, and place nontrivial constraints on 3d gauge theories~\cite{Albayrak2021-td,Chester2016-tp}. These results are obtained by combining SDP constraints involving a variety of ``light" correlators that relate the OPEs of relevant and marginal operators in the theory. In the conformal block expansion of these correlators, one observes that the OPE coefficients of light operators are large compared to the corresponding coefficients for heavy operators. When this is the case, we say that the correlator is ``dominated" by light operators. Moreso, contributions from heavy operators are further suppressed by the exponential decay of the conformal block around the crossing symmetric configuration of $z = \bar{z} = 1/2$. The result of this fact is that the OPE decomposition of light correlators may be effectively truncated to the low-lying spectrum, so that the parameter space subject to optimization is sufficiently small and the numerics are tractable. 

A class of observables that remains somewhat elusive to standard numerical bootstrap treatment are four-point correlators which involve some number of irrelevant operators. One reason for this is that correlation functions involving operators with scaling dimension much larger than the unitarity bound tend to receive important contributions from a large number of exchanged operators with scaling dimensions of the same order. This makes the parameter space subject to optimization much larger than can be effectively analyzed numerically, with extremal functionals from SDP converging very slowly for larger values of $\Delta$. If one were able to overcome these difficulties, correlators of irrelevant operators would give us better access to the large scaling dimension data of the CFT. These data can give us insights into the mechanics of strongly-interacting multiparticle and black hole states in the dual gravitational bulk -- unraveling the mysteries of which is a crucial goal in the study of quantum gravity.

To better clarify our observables of interest in the context of holography, consider a scalar operator $\phi$ in a $d$-dimensional boundary CFT. The scaling dimension of this operator is related to the mass of the corresponding bulk field by
\ba
\Df (\Df - d) = m^2,
\ea
where we work in units where the AdS curvature $R=1$. Taking $\Df$ larger implies a larger mass, but how do we quantify ``heavy"? When the boundary CFT has a conserved stress tensor, there is a finite central charge $C_T$ to which we can compare the scaling dimensions of boundary operators.\footnote{The squared OPE coefficient describing the three point coupling of two identical scalar operators $\phi$ to the stress tensor $T$ is $\propto\frac{\Df^2}{C_T}$ where $C_T$ is the central charge of the theory.} In the case of $\Df \gtrsim \sqrt{C_T}$, we can justly consider the operator ``heavy," and its insertion on the boundary distorts the AdS metric~\cite{Mishra2024-mr}. The exact bulk description of these operators is theory dependent, and they may be dual to strings, branes, or black holes emerging from the asymptotic boundary. Computing boundary correlators of these operators holographically requires corrections from the presence of these extended surfaces in the bulk. For $\Df \lesssim \sqrt{C_T}$, the boundary insertions are more generally viewed as insertions of massive particles, Kaluza-Klein (KK) modes, or perhaps de-localized ``blobs"~\cite{Fardelli2024-ra,Abajian2023-xw} (depending on the presence of a large $N$ parameter) and can be in principle computed with Witten diagrams. In this work, we will generally refer to any correlator with external scaling dimension $\Df \gg \frac{d-2}{2}$ as heavy, and we will refer to the $\Df \to \infty$ limit as the heavy limit. 

The majority of extant literature on heavy dynamics focuses on the case of heavy-heavy-light-light correlators, see e.g.~\cite{Abajian2023-xw, Grabovsky2024-bg}. These correlators are amenable to a variety of holographic approaches where the heavy states source a background geometry in AdS space and light operators are approximated as ``probes" which travel along geodesic paths in the deformed spacetime. These correlators can also be related to the two-point functions of light operators in a CFT at finite temperature, which describes the dynamics of a light operator scattering off an AdS black hole with the same Hawking temperature~\cite{Iliesiu2018-od}.

These approaches fail in the case of heavy-heavy-heavy-heavy correlators where each of the operators is both sourcing and backreacting off of each other's geometry. Not only is this problem difficult within a known theory, but attempting to study them from the bootstrap perspective seems similarly intractable as there is very little known about how to effectively truncate the parameter space that characterizes them. Unlike light correlators whose behavior is well approximated by a finite number of quantum numbers describing the low-lying spectrum, there is no such immediate ``microscopic" description that captures the physics of heavy dynamics where the OPE is dominated by a large number of similarly heavy operators.

Finite temperature calculations have been used to derive high-energy asymptotics of CFT data, including heavy-heavy-heavy OPE coefficients and the asymptotic density of states for a general dimension CFT as a function of scaling dimension and spin~\cite{Benjamin2023-qo}. These techniques work by describing thermal correlators as local operators coupled to background fields governed by a local ``thermal" effective action on compact geometries. On a $S^1_\beta \times S^{d-1}$ torus, this effective action describes a thermal partition function with inverse temperature $\beta$ and spin fugacity $\vec{\Omega}$, which parametrizes ``twists" of $S^{d-1}$ along the thermal circle. 

\begin{figure}[tbp]
    \centering
    \includegraphics[width=.8\linewidth]{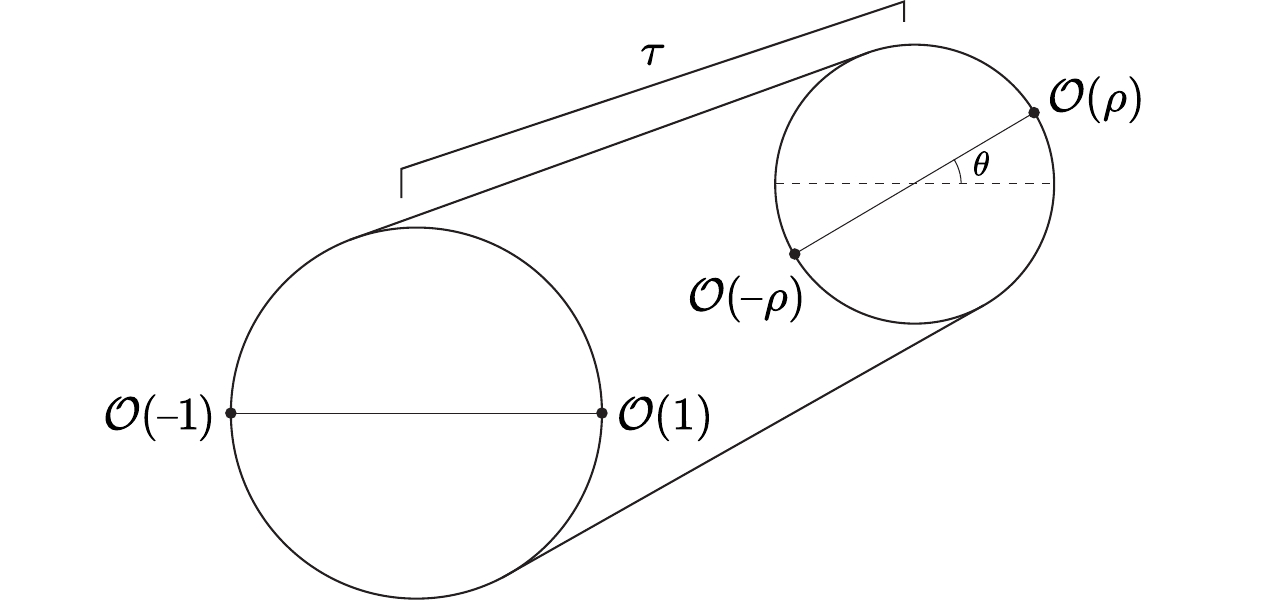}
    \caption{Operators in a four-point correlator placed on a cylinder. When the operators are sufficiently heavy, the conformal block resembles a Gibbs measure with inverse temperature $\beta \sim -\tau$ and 1d spin fugacity $\vec{\Omega} \sim \theta$. Macroscopic observables are produced by generating ``twists" and ``pulls" of the operator pairs. Under the integral transformation $\mathcal{T}$ introduced in section~\ref{sec:operators}, this relation becomes exact in 1, 2, and 4 spatial dimensions. }
    \label{fig:cyl}
\end{figure}

When a four-point correlator of identical scalars is dominated by operators with large scaling dimension, the conformal block decomposition resembles a thermal partition function for the subset of states that show up in the OPE of the external operators. In this limit, the effective inverse temperature is controlled by the separation of operator pairs along the cylinder $-\tau$ (see figure~\ref{fig:cyl}), a 1-dimensional spin fugacity is controlled by the angular separation $\theta$ of operator pairs along the cylinder, and the OPE coefficients resemble state degeneracy factors. This leads us to consider a similar space of observables to characterize heavy correlators as we do more general thermal systems, where standard macroscopic observables such as average energy and total angular momentum can be computed by applying appropriate functionals to the correlator.

In this paper, we will study correlators of identical scalars with $\Df \gg \frac{d-2}{2}$ using a three-fold approach. First, in section~\ref{sec:operators} we define and construct Riemmann-Liouville type fractional differential operators in 1, 2, and 4 dimensions that extract the principal series eigenvalues from a conformal block. These operators resolve the 8-fold degeneracy in the eigenspace of the quadratic Casimir of the conformal group into four subspaces related by a discrete ``rotation" symmetry of the Casimir eigenvalue. Additionally, we construct the analogous operators in general dimension that extract these eigenvalues from the asymptotic conformal block at large scaling dimension. When applied to a CFT correlator of identical scalars, these ``principal series operators" compute global averages over CFT data in a given kinematic regime, weighted by powers of the quantum numbers of scaling dimension $\Delta$ and total angular momentum $J_2$. These global averages are mathematically understood as classical moments in a double Stieltjes problem, and we prove that an infinite sequence of these moments can be used to uniquely recover the CFT data which decomposes the correlator.

Second, in section~\ref{sec:moments} we directly use constraints from crossing and unitarity to derive bounds and relations between these moments, focusing on the ``heavy" limit of $\Df \to \infty$. Previous studies have analyzed constraints from crossing in this limit~\cite{Kim2015-sg} and have derived leading-order relations between moments in scaling dimension~\cite{Paulos2016-mh}. We further extend these results by deriving subleading relations between moments, including those which involve some power of total angular momentum. We then combine these relations with positivity constraints from unitarity to derive a leading bound on the covariance of the quantum numbers $J_2$ and $\Delta$ in any crossing-symmetric OPE, as well as two-sided bounds on the leading behavior of moments in scaling dimension.

In section~\ref{sec:saddles} we re-sum the moment sequences that saturate the leading bounds to derive extremal saddle-point solutions to the crossing equations in the heavy limit. This leads us to our last fold, where we relate these solutions to particular limits of correlators in a generalized free field (GFF) theory, and show that saddle points in the OPE distribution correspond to a decomposition into higher-spin (HS) conformal blocks which organize contributions to the GFF OPE by operators involving a fixed number of elementary scalar fields. These HS conformal blocks are known to provide a tractable finite basis for correlators with weakly broken higher-spin symmetry~\cite{Alday2016-kz}. Further, we show that the derived measures we obtain by matching only the second moment of these HS conformal blocks satisfy the properties of a function which, up to a determined factor, interpolates the weights of operators in the OPE.\footnote{The weight of an operator $\cO$ in a $\phi \times \phi$ OPE is given by $\lambda^2_{\phi \phi \cO} G_\cO(z,\bar{z})$ and is thus dependent on the kinematics of the correlator.} We show that these weight-interpolating functions (WIFs) provide quantitative predictions of OPE coefficients as a continuous function of scaling dimension for correlators of sufficiently large external scaling dimension in a variety of perturbative examples. A corollary to this observation is that the weights of operator families of fixed length become distributed along Gaussian distributions, therefore reducing the space of variables that describes them to their first few moments. We conclude with a discussion of our results in section~\ref{sec:discussion}.

\section{Principal series operators} 
\label{sec:operators}

Let us begin by considering the conformal group, with generators 
\ba
\left[M_{\mu\nu},P_\rho\right] &= \delta_{\nu\rho}P_\mu - \delta_{\mu\rho}P_\nu,\\
\left[M_{\mu\nu},K_\rho\right] &= \delta_{\nu\rho}K_\mu - \delta_{\mu\rho}K_\nu,\\
\left[M_{\mu\nu},M_{\rho \sigma}\right] &= \delta_{\nu\rho}M_{\mu\sigma}-\delta_{\mu\rho}M_{\nu\sigma}+\delta_{\nu\sigma}M_{\rho\mu}-\delta_{\mu\sigma}M_{\rho\nu},\\
\left[D,P_\mu\right]&=P_\mu,\\
\left[D,K_\mu\right]&=-K_\mu,\\
\left[K_\mu,P_\nu\right]&=2\delta_{\mu\nu}D-2M_{\mu\nu}.
\ea
The $d$-dimensional Lorentzian conformal algebra is isomorphic to the algebra of $SO(d,2)$, with its generators $L_{AB}$ identified as
\ba
L_{\mu\nu}&=M_{\mu\nu},\\
L_{-1,0} &= D,\\
L_{0,\mu} &= \frac 1 2 (P_\mu+K_\mu),\\
L_{-1,\mu}&= \frac 1 2 (P_\mu-K_\mu).
\ea
Here, $L_{-1,0}$ and $L_{0,\mu}$ are non-compact and generate dilatations and longitudinal Lorentz boosts respectively~\cite{agarwal1}. These generators give rise to unitary principal series representations $\mathcal{P}_{\Delta,\ell,\lambda}$ labeled by continuous weights $\Delta = \frac{d}{2} + i s$, $\ell = -\frac{d-2}{2}+ i q$ for $s,q \in \mathbb{R}$, and an irreducible representation $\lambda$ of $SO(d-2)$. Together, the pair $(\ell,\lambda)$ specifies a weight of $SO(d)$, and $\ell$ is the length of the first row in its Young Tableaux diagram~\cite{Kravchuk_2018}. To simplify our discussion, we will take $\lambda$ to be the trivial representation so that $\mathcal{P}_{\Delta,\ell,\lambda}$ is a rank$-\ell$ traceless symmetric tensor, and suppress the $\lambda$ label $\mathcal{P}_{\Delta,\ell,\lambda} \to \mathcal{P}_{\Delta,\ell}$. 

The principal series representation $\mathcal{P}_{\Delta,\ell}$ is an eigenvector of the quadratic Casimir of the conformal group $C_2 = \frac{1}{2}L^{AB}L_{AB}$, with eigenvalue
\ba
C_2 (\mathcal{P}_{\Delta,J}) = \Delta(\Delta-d) + \ell(\ell+d-2).
\ea
This eigenvalue has a discrete symmetry group isomorphic to the dihedral group $D_8$, which includes three $\mathbb{Z}_2$ subgroups given by the actions
\ba
z_1: \;\Delta \leftrightarrow d- \Delta, \;  z_2: \;\Delta \leftrightarrow 1-\ell , \; z_3:\;\ell \leftrightarrow 2-d-\ell.
\ea
Rewriting $r = z_1 z_2$ and $s = z_3$ (or alternatively $r= z_3 z_2$ and $s = z_1$) we see that $r$ generates rotations and $s$ generates reflections of the square, giving the standard $D_8$ group presentation
\ba
\langle r, s | r^4 = s^2 = (r s)^2 = 1\rangle.
\ea
Since $|D_8| = 8$, the eigenspace of $C_2$ is 8-fold degenerate and its eigenbasis is obtained by applying group elements of $D_8$ to $\mathcal{P}_{\Delta,\ell}$. The resulting basis is given by
\ba
\begin{array}{c|c}
\hline
\mathrm{Element \;of\; } D_8 \; & C_2\; \mathrm{eigenvector}\\
\hline
1 \; & \mathcal{P}_{\Delta,\ell}\\
r \; & \mathcal{P}_{1-\ell,-d+\Delta+1}\\
r^2 \; & \mathcal{P}_{d-\Delta,-d-\ell+2}\\
r^3 \; & \mathcal{P}_{d+\ell-1,1-\Delta}\\
s \; & \mathcal{P}_{\Delta,-d-\ell+2}\\
r \circ s \; & \mathcal{P}_{1-\ell,1-\Delta}\\
r^2 \circ s \; & \mathcal{P}_{d-\Delta,\ell}\\
r ^3\circ s \; & \mathcal{P}_{d+\ell-1,-d+\Delta+1}\\
\hline
\end{array}
\ea

In this section, we will explicitly construct additional operators $\Omega_+$ and $\Omega^2_-$ in $d=1,2$ and $4$ which extract the principal series eigenvalues, $is = \Delta - \frac{d}{2}$ and $-q^2 = \left( \ell+\frac{d-2}{2}\right)^2$ respectively, from eigenvectors of $C_2$. These operators decompose the quadratic Casimir as
\ba
\frac{1}{2}\left(\Omega_+^2+\Omega_-^2-\left(\frac{d}{2}\right)^2-\left(\frac{d-2}{2}\right)^2 \right) = C_2
\ea
and allow us to resolve the 8-fold degeneracy of the $C_2$ eigenspace into 4 independent 2-fold degenerate subspaces given by
\ba
\begin{array}{c|c|c|c}
\hline
\mathrm{Element\; of\;} D_8 \, &\mathrm{Subspace}\, & \Omega_+\, & \Omega^2_-\,\\ \hline 
1,s \, & \mathcal{P}_{\Delta,\ell},\mathcal{P}_{\Delta,2-d-\ell} \, & \Delta - \frac{d}{2} \, & \frac{1}{4} (d+2 \ell-2)^2 \, \\
 r,r \circ s \, &\mathcal{P}_{1-\ell,1-d+\Delta}, \mathcal{P}_{1-\ell,1-\Delta} \, &   1-\frac{d}{2}-\ell \, & \frac{1}{4} (d-2 \Delta)^2 \, \\

 r^2,r^2\circ s \, & \mathcal{P}_{d-\Delta,2-d-\ell},\mathcal{P}_{d-\Delta,\ell} \, &   \frac{d}{2}-\Delta \, & \frac{1}{4} (d+2 \ell-2)^2 \, \\

r^3, r^3 \circ s \, & \mathcal{P}_{d-1+\ell,1-\Delta},\mathcal{P}_{d-1+\ell,1-d+\Delta} \, &   \ell+\frac{d}{2}-1 \, & \frac{1}{4} (d-2 \Delta)^2 \, \\
\hline
\end{array}
\ea
Elements within a subspace are related by precomposing a rotation with $s$, and the remaining degenerate subspaces are related by the rotation $r$. 

Concretely, we will be constructing $\Omega_\pm$ by studying the actions of integral transforms on conformal blocks $G_{\Delta,\ell}$. Conformal blocks can be schematically written as
\ba
G_{\Delta,\ell} \sim \frac{\langle \phi_1(x_1) \phi_2(x_2) \mathcal{O} \rangle \langle \mathcal{O}\phi_3(x_3) \phi_4(x_4)  \rangle}{ \langle \mathcal{O}\mathcal{O}\rangle},
\ea
where $\phi_i(x_i)$ are ``external" local scalar operators at marked points $x_i \in S^d$, and $\mathcal{O}$ is an ``exchanged" local primary operator with quantum numbers $\Delta,\ell$. Conformal blocks are group harmonics for a conformal correlator of the form $\langle \phi_1(x_1) \phi_2(x_2)\phi_3(x_3) \phi_4(x_4)  \rangle $, repackaging contributions of irreducible representations with lowest weight vector $\mathcal{O}$ to the correlator. 

When the external scalar operators are identical, the conformal block satisfies the second-order differential equation
\ba
\hat{C}_2 G_{\Delta,\ell} = \frac{1}{2}\left( \Delta (\Delta-d) + \ell (\ell+d-2)\right) G_{\Delta,\ell},
\ea
where
\ba
\hat{C}_2 = \mathcal{D}_z+\mathcal{D}_{\bar{z}}+(d-2)\frac{z\bar{z}}{z-\bar{z}}((1-z)\partial_z-(1-\bar{z})\partial_{\bar{z}})
\ea
and
\ba
\mathcal{D}_z = z^2(1-z)\partial^2_z-z^2\partial_z.
\ea
The ``Dolan-Osborn" coordinates $z,\bar{z}$ are related to the standard conformal invariant cross ratios as $z \bar{z} = u = \frac{x^2_{12}x^2_{34}}{x^2_{13}x^2_{24}}$ and $(1-z)(1- \bar{z}) = v = \frac{x^2_{14}x^2_{23}}{x^2_{13}x^2_{24}}$. In $d = 1, 2,$ and $4$, exact solutions to the differential equation are given by~\cite{Dolan:2000ut,Dolan:2003hv}
\ba
G^{(1)}_\Delta(z) &= k_\Delta(z) = z^\Delta {}_2 F_1\left(\Delta,\Delta,2\Delta; z\right),\\
G^{(2)}_{\Delta,\ell}(z,\bar{z}) &= \frac{(-1)^\ell}{2^\ell (\delta _{\ell,0}+1)} \left( k_{\frac{\Delta+\ell}{2}}(z)k_{\frac{\Delta-\ell}{2}}(\bar{z})  + z\leftrightarrow \bar{z}\right),\\
G^{(4)}_{\Delta,\ell}(z,\bar{z}) &= \frac{(-1)^\ell}{2^\ell }\frac{z\bar{z}}{z-\bar{z}} \left( k_{\frac{\Delta+\ell}{2}}(z)k_{\frac{\Delta-\ell-2}{2}}(\bar{z})  - z\leftrightarrow \bar{z}\right).
\ea

In general dimension, conformal blocks also admit the radial representation~\cite{Hogervorst:2013sma}
\ba
G_{\Delta,\ell}(r,\eta) =(4 r)^\Delta h_{\Delta,\ell}(r,\eta),
\ea
where 
\ba
h_{\Delta,\ell}(r,\eta) = h^{\infty}_\ell(r,\eta) + h^{\mathrm{null}}_{\Delta,\ell}(r,\eta)
\ea
is the regulated conformal block, and 
\ba
r &= \sqrt{\rho \bar{\rho}} = \frac{ \sqrt{z \bar{z}}}{\left(\sqrt{1-z}+1\right) \left(\sqrt{1-\bar{z}}+1\right)},\\ 
\eta &= \frac{\rho + \bar{\rho}}{2 \sqrt{\rho \bar{\rho}}}=\frac{1-\sqrt{(1-\bar{z} ) (1- z)}}{\sqrt{z \bar{z} }}
,\\ \rho &= \frac{1-\sqrt{1-z}}{1+\sqrt{1-z}}
\ea
are radial coordinates.

Here, $h_{\Delta,\ell}(r,\eta)$ is a meromorphic function in $\Delta$ with $h^{\mathrm{null}}_{\Delta,\ell}(r,\eta)$ containing a series of poles in $(\Delta,\ell)$ below the unitarity bound associated with zero-norm vectors~\cite{Kos:2013tga, Penedones_2016}. Explicitly,
\ba
h^\infty_\ell(r,\eta) = \mathcal{N}_{d,\ell} \frac{\left(1-r^2\right)^{1-\frac{d}{2}} }{\sqrt{r^2-2 \eta  r+1} \sqrt{r^2+2 \eta  r+1}}C_\ell^{\left(\frac{d-2}{2}\right)}(\eta )
\label{asymregblock}
\ea
and
\ba
h^{\mathrm{null}}_{\Delta,\ell}(r,\eta) = \sum_A \frac{R_A}{\Delta-\Delta_A^*} (4r)^{n_A} h_{\Delta_A^*+n_A,\ell_A}(r,\eta),
\ea
with $\mathcal{N}_{d,\ell} = \frac{\ell!}{(-2)^\ell \left(\frac{d}{2}-1\right)_\ell}$ a Gegenbauer normalization factor, and $A$ indexes the infinite set of null states. While $h^{\mathrm{null}}_{\Delta,\ell}(r,\eta)$ is not known in closed form, it can be computed recursively order-by-order in powers of $r$. Due to the presence of the poles, the contribution of the null states is suppressed as $\Delta \to \infty$, allowing us to determine the asymptotics of the conformal blocks for $\Delta \gg \frac{d-2}{2}$ in general dimension as\footnote{Considering asymptotics of $x \to \infty$, we say $f(x) = O(g(x))$ if $\exists$ $M>0$ such that $|f(x)| \leq M |g(x)|$ for all sufficiently large $x$.}
\ba
G_{\Delta, \ell}(r,\eta) = (4 r)^\Delta \left( h_\ell^\infty(r,\eta) + O\left(\frac{1}{\Delta} \right) \right).
\label{confblockasym}
\ea

\subsection{Exact operators}
In~\cite{song2023compactform3dconformal}, a modified Riemann–Liouville fractional derivative was introduced with the following transformation property on $d=1$ conformal blocks
\ba
T_z^{\frac{1}{2}} k_{\Delta}(z) = \frac{4^{\Delta-1/2} \Gamma(\Delta-1/2)}{\Gamma(\Delta)} \rho^{\Delta-1/2},
\label{powerlawtrans}
\ea
where
\ba
T^\delta_z f(z)= \frac{\partial^{-\delta}}{\partial^{-\delta}(-1/z)} f(z) =\frac{z^{-\delta}}{\Gamma(\delta)} \int_0^1 dt \; f( z t) (1-t)^{\delta-1}t^{-\delta-1}.
\ea
The prefactor of (\ref{powerlawtrans}) vanishes for $\Delta = 0$, so this transformation acts as zero on the conformal block associated with the exchange of the identity operator. By conjugating an infinitesimal rescaling of $\rho$ by this transformation, we can construct an operation which extracts the principal series eigenvalue 
\ba
T_z^{-\frac{1}{2}} \rho \partial_\rho T_z^{\frac{1}{2}}k_{\Delta}(z) = \left(\Delta-\frac{1}{2}\right) k_{\Delta}(z).
\ea
In $d=2$ and 4, the conformal block factorizes up to a power of $\frac{z-\bar{z}}{z\bar{z}}$ into a $z\leftrightarrow\bar{z}$ symmetric sum of products of $d=1$ conformal blocks, so it is possible to construct an analogous $\mathcal{T}$ operator that transforms a conformal block into a sum of power-laws in $\rho,\bar{\rho}$. 

We denote this transformation
\ba
\mathcal{T} = T^{1/2}_z T^{1/2}_{\bar{z}}\left(\frac{z-\bar{z}}{z\bar{z}}\right)^{\frac{d-2}{2}}
\ea
and define
\ba
\Omega_\pm = \mathcal{T}^{-1}\left( \rho \partial_\rho \pm \bar{\rho} \partial_{\bar{\rho}}  \right) \mathcal{T}.
\ea
It is easy to check that the $\Omega_\pm$ operators satisfy
\ba
\Omega_+ G_{\Delta,\ell}(z,\bar{z}) = \left(\Delta-\frac{d}{2}\right) G_{\Delta,\ell}(z,\bar{z}),
\ea
\ba
\Omega_-^2 G_{\Delta,\ell}(z,\bar{z}) = \left(\ell+\frac{d-2}{2}\right)^2 G_{\Delta,\ell}(z,\bar{z}),
\ea
in 2 and 4 dimensions, where $\Omega_-$ must be applied twice to extract an eigenvalue from the conformal block due to its antisymmetry under $z \leftrightarrow \bar{z}$.\footnote{When $\Omega_-$ is applied once, the conformal block is transformed to a chirally antisymmetric block which flips signs under $z \leftrightarrow \bar{z}$.} Indeed, it is this $\mathbb{Z}_2$ symmetry of the quadratic Casimir under $z \leftrightarrow \bar{z}$ that renders $\Omega_\pm$ unable to resolve the remaining 2-fold degeneracy of the $C_2$ eigenspace.

\subsection{Asymptotic operators}

In addition to being able to construct the $\Omega$-operators exactly in $1,2,$ and 4 dimensions, we can also use the known form of the conformal block at large $\Delta$ to construct analogous asymptotic operators in general dimension, which we will call $\tilde{\Omega}_{\pm}$.

Upon a Weyl transformation of $\mathbb{C} \to S^1 \times  \mathbb{R} $, we introduce the cylinder coordinates
\ba
\tau &= \log(\sqrt{\rho \bar{\rho}}),\\
i\theta & = \log\left( \sqrt{\frac{\rho}{\bar{\rho}}}\right),
\label{cylindercoords}
\ea
so that $ \Omega_+ = \mathcal{T}^{-1} \partial_\tau\mathcal{T} $ and $\Omega_- =- i \mathcal{T}^{-1} \partial_\theta \mathcal{T}$. Since $\mathcal{T}^{-1} \mathcal{T}=1$, we have
\ba
\Omega_+ = \mathcal{T}^{-1}[\partial_\tau,\mathcal{T} ] + \partial_\tau
\label{opgenform}
\ea

Applying this operator to the radial form for the conformal block and solving for the action of $\mathcal{T}^{-1}[\partial_\tau,\mathcal{T} ]$ gives
\ba
 \mathcal{T}^{-1}  [\partial_{\tau} ,\mathcal{T}] G_{\Delta,\ell}(\tau,\theta)=\left(-\frac{\partial_\tau h_{\Delta,\ell}(\tau,\theta)}{ h_{\Delta,\ell}(\tau,\theta)}-d/2\right)  G_{\Delta,\ell}(\tau,\theta).
\ea
We can now use the known form of $h^{\infty}_\ell(\tau,\theta)$ to find
\ba
 \mathcal{T}^{-1}  [\partial_\tau ,\mathcal{T}] G_{\Delta,\ell}(\tau,\theta) =\left( \frac{\coth (\tau ) ((d-2) \cos (2 \theta )-d \cosh (2 \tau )+2)}{2 (\cos (2 \theta )-\cosh (2 \tau ))} + O\left(\frac{1}{\Delta} \right) \right)G_{\Delta,\ell}(\tau,\theta),
\ea
which acts as multiplication by a $\Delta$-independent function on nonidentity blocks, and zero on the identity, so that the asymptotic result is consistent with the exact result. 
Combining with (\ref{opgenform}), we find
\ba
\tilde{\Omega}_+ = \frac{\coth (\tau ) ((d-2) \cos (2 \theta )-d \cosh (2 \tau )+2)}{2 (\cos (2 \theta )-\cosh (2 \tau ))} + \partial_\tau,
\ea
which satisfies $\tilde{\Omega}_+ G_{\Delta,\ell}(\tau,\theta) \sim \left(\Delta-d/2\right) G_{\Delta,\ell}$ as $\Delta \to \infty$.

We can attempt a similar procedure to compute $\tilde{\Omega}_-$, however one quickly finds that  $\mathcal{T}^{-1} [ \partial^2_\theta,\mathcal{T}]$ does not act as a function independent of $\ell$ on the conformal block, telling us there is additional mixing of differential operators when $\tilde{\Omega}_-$ is applied twice. Furthermore, since a conformal block is not an eigenfunction of $\Omega_-$, we cannot use an analogous operator equation to compute $\mathcal{T}^{-1}[\partial_\theta,\mathcal{T}]$ alone. 

Instead, we will start by constructing an operator that acts as $\tilde{J}_2 h_\ell^\infty =\ell(\ell+d-2) h_\ell^\infty$ by using the property of the Gegenbauer polynomial
\ba
\mathcal{J} C^{\frac{d-2}{2}}_\ell(\eta) = - \ell(\ell+d-2)C^{\left(\frac{d-2}{2}\right)}_\ell(\eta)
\ea
with
\ba
\mathcal{J} = (1-\eta^2)\partial^2_\eta +(1-d)\eta \partial_\eta.
\ea

We can then construct $\tilde{J}_2$ by dressing $\mathcal{J}$ with a term which subtracts off the remaining commutator
\ba
\left[\mathcal{J},\frac{\left(1-r^2\right)^{1-\frac{d}{2}} }{\sqrt{r^2-2 \eta  r+1} \sqrt{r^2+2 \eta  r+1}}\right]C_\ell^{\left(\frac{d-2}{2}\right)}(\eta ).
\ea
Since $\mathcal{J}$ is second order, the commutator is a first-order differential operator which we compute directly, giving
\ba
\tilde{J}_2 =  \frac{16 (d-1) \eta ^4 r^4-4 d \eta ^2 \left(r^3+r\right)^2+4 \left(r^3+r\right)^2}{\left(r^4+\left(2-4 \eta ^2\right) r^2+1\right)^2}-\frac{8 \eta  \left(\eta ^2-1\right) r^2}{r^4+\left(2-4 \eta ^2\right) r^2+1}\partial_\eta - \mathcal{J},
\ea
which satisfies $\tilde{J}_2 G_{\Delta,\ell} \sim \ell(\ell+d-2)G_{\Delta,\ell}$ in the limit of $\Delta \to \infty$.

Assembling the $\Omega$-operators with the appropriate dimension-dependent shifts then gives the final result
\ba
\tilde{\Omega}_+ =&\, \frac{\left(r^2+1\right) \left(-4 (d-2) \eta ^2 r^2+d \left(r^2+1\right)^2-8 r^2\right)}{2 \left(r^2-1\right) \left(r^4+\left(2-4 \eta ^2\right) r^2+1\right)}+r \partial_r,\\
\tilde{\Omega}^2_- =&\,  \frac{16 (d-1) \eta ^4 r^4-4 d \eta ^2 \left(r^3+r\right)^2+4 \left(r^3+r\right)^2}{\left(r^4+\left(2-4 \eta ^2\right) r^2+1\right)^2} \\
&-
\left(\frac{8 \eta  \left(\eta ^2-1\right) r^2}{r^4+\left(2-4 \eta ^2\right) r^2+1} + (1-d)\eta\right)\partial_\eta 
-  (1-\eta^2)\partial^2_\eta 
+ \left(\frac{d-2}{2}\right)^2.
\ea

\section{Moments of the OPE}
\label{sec:moments}

In this section we will shift gears and review some of the mathematics of classical moment problems, with the goal of reinterpreting CFT correlators in this language. Crossing symmetry will then impose constraints on the moments of CFT correlators, leading to nontrivial bounds on these moments.

\subsection{Review of classical moment problems}
First, we will briefly review some key topics and results for the classical moment problem. For an in-depth discussion in mathematics literature, see the standard texts by Akhiezer~\cite{alma99671493502466}, Shohat \& Tamarkin~\cite{Shohat1943ThePO}, and Schm\"udgen~\cite{gradtextsmomentproblem}.

A classical moment problem studies the moment map which takes a positive distribution function $f$ on $X \subseteq \mathbb{R}$ to the sequence of moments $m_\bullet = (m_n)_{n \geq0}$ given by
\ba
m_n = L_f[X^n] = \int_X x^n f(x)dx .
\ea
Given such a sequence, we would like to determine the following: 1) if such a positive measure $f$ exists, 2) if the moment sequence uniquely determines $f$. If these two conditions are satisfied, then the moment sequence is said to be ``determinant." For many determinant moment sequences, there exists a moment-generating function (MGF) $M_X(t) = L_f[e^{X t}]$ which is bounded in some interval $t \in ( -t_0,t_0)$ for $t_0>0$ and satisfies $\partial_t^n M_X(t)|_{t=0} = m_n$ for all $n\geq0$. Moreso, the measure $f$ can be uniquely recovered from $M_X(t)$ by applying the inverse Laplace transform
\ba
f(x) = \int_{\gamma-i\infty}^{\gamma+i\infty} \frac{dt}{2\pi i} e^{-x t} M_X(t).
\ea

\paragraph{Hamburger moment problem}
The classical Hamburger moment problem is the case when $X = \mathbb{R}$. If $f$ is a positive measure, then $L_f[  P(X)^2 ] \geq 0$ for all polynomials $P(X) \in \mathbb{R}[X]$. If we write an $n$-th degree polynomial as $P^n(X) = \sum^{n}_{k} a_k X^k $, then the previous condition is equivalent to $\sum^n_{j,k} a_j a_k m^{j+k} =\mathbf{a}^\top H_n^{(0)} \mathbf{a} \geq 0 $ for all $\mathbf{a} \in \mathbb{R}^n$, where
\ba
H^{(0)}_n=\left[\begin{matrix}
m_0 & m_1 & m_2 & \cdots & m_{n}    \\
m_1 & m_2 & m_3 & \cdots & m_{n+1} \\
m_2& m_3 & m_4 & \cdots & m_{n+2} \\
\vdots & \vdots & \vdots & \ddots & \vdots \\
m_{n} & m_{n+1} & m_{n+2} & \cdots & m_{2n}
\end{matrix}\right]
\ea
is a positive semi-definite symmetric ``Hankel matrix" of moments. Thus, given a moment sequence $m_\bullet$, a functional $L_f$ exists if and only if
\ba
H^{(0)}_n  \succeq 0 \quad\quad \forall n \in \mathbb{N}.
\ea

Sylvester's criterion for the positivity of symmetric matrices states that this condition is satisfied if the determinants of all leading minors are non-negative. Additionally, the set of positive semi-definite matrices $\mathcal{C}$ forms a closed convex cone; this means that for all $C_1 ,C_2 \in \mathcal{C}$ and $\alpha,\beta>0$ we have $\alpha C_1 + \beta C_2 \in \mathcal{C} $. Thus, we can view moment sequences that satisfy Hankel matrix positivity as living in a convex subset of all positive real sequences called the moment cone~\cite{schmudgen2020lecturesmomentproblem}. We can projectivize this space by normalizing all moments by $m_0$. Under this projectivization, the convexity of the moment cone can be understood as the following: given any two normalized moment sequences $\left(m_n/m_0\right)_{(0)}$, $\left(m_n/m_0\right)_{(1)}$ in the moment cone, the sequence $\left(m_n/m_0\right)_{(2)} = \lambda \left(m_n/m_0\right)_{(0)} + (1 - \lambda)\left(m_n/m_0\right)_{(1)} $ for all $\lambda \in (0,1)$ is also in the moment cone.

A sufficient criterion for uniqueness is given by Carleman's condition for the Hamburger problem, which states that a moment sequence $m_\bullet$ is determinant if
\ba
\sum_{n\geq1} m_{2n}^{-1/(2n)} = +\infty.
\label{carlemanshamburger}
\ea

\paragraph{Stieltjes moment problem}
The classical Stieltjes moment problem is the case when $X  = [0,\infty)$. Since this requires both the support and measure to be positive, we have the stronger condition that $L_f[ X^k P(X)^2 ] \geq 0$ for all polynomials $P(X) \in \mathbb{R}[X]$ and $k\geq 0$. This condition gives rise to the following criteria for existence: let 
\ba
H^{(1)}_n=\left[\begin{matrix}
m_1 & m_2 & m_3 & \cdots & m_{n+1}    \\
m_2 & m_3 & m_4 & \cdots & m_{n+2} \\
m_3& m_4 & m_5 & \cdots & m_{n+3} \\
\vdots & \vdots & \vdots & \ddots & \vdots \\
m_{n+1} & m_{n+2} & m_{n+3} & \cdots & m_{2n+1}
\end{matrix}\right]
\ea
denote the shifted Hankel matrix. If
\ba
H^{(0)}_n \succeq 0, \quad H^{(1)}_n \succeq 0 \quad \quad \forall n \in \mathbb{N},
\ea
then the functional $L_f$ exists, with $\mathrm{supp}(f) \subset [0,\infty)$.

The analogous Carleman's condition for the Stieltjes problem states that a moment sequence $m_\bullet$ is determinant if
\ba
\sum_{n\geq1} m_n^{-1/(2n)} =+ \infty.
\ea

\paragraph{Double moment problem}
The Hamburger and Stieltjes moment problems can be readily generalized to measures with higher dimensional support. Let $f: X\times Y \to \mathbb{R}$ be a positive distribution function over the product space $X \times Y$. Denote
\ba
m_{p,q} = \int_X \int_Y dx\;dy\;  x^p y^q f(x,y)
\ea
as the corresponding double-moments of the measure. 

We construct the generalized Hankel matrix
\ba
H^{(j,k)}_{n} =\left[\begin{matrix}
m_{j,k} & m_{j,1+k} & \cdots & m_{j,n+k}   \\
m_{1+j,k} & m_{1+j,1+k}  & \cdots & m_{1+j,n+k} \\
\vdots & \vdots  & \ddots & \vdots \\
m_{n+j,k} & m_{n+j,1+k} & \cdots & m_{n+j,n+k}
\end{matrix}\right].
\ea
Focusing on the Stieltjes case, given a double moment sequence $m_\bullet = \{ m_{p,q} \}_{p,q \geq 0}$ a positive functional $L_f$ exists if and only if
\ba
H^{(1,0)}_{n} \succeq 0,\quad H^{(0,1)}_{n} \succeq 0 \quad \quad \forall n \in\mathbb{N}.
\ea
Additionally, the generalized Carleman's condition states that such a functional $L_f$ is uniquely determined by the double-moment sequence if
\ba
\sum_{n \geq 1} m_{n,0}^{-1/(2n)} = + \infty, \quad\sum_{n \geq 1} m_{0,n}^{-1/(2n)} = + \infty.
\ea

In other words, a measure $f$ on the product space $X \times Y$ is determined by its double-moment sequence if the reduced measures $f(x) = \int_Ydy \; f(x,y)$, $f(y) = \int_X dx \; f(x,y)$ are determined by their respective single-moment sequences.

\subsection{Four-point correlators in CFT}

Correlation functions of four identical scalar operators, $\phi$, in a unitary CFT are closely related to a classical moment generating function for a positive measure describing the $\phi \times \phi$ OPE. In this work, we study a measure over scaling dimension $\Delta$ and total angular momentum $J_2 \equiv \ell(\ell+d-2)$ that arises naturally from the conformal block decomposition of a correlator.

Let us fix some kinematics $(z,\bar{z}) \to (z^\star,\bar{z}^\star)$ where the conformal block $G_{\Delta,\ell}(z^\star,\bar{z}^\star)$ is non-negative for all $\Delta,\ell$. The existence of such kinematics arise from the property of reflection positivity in a unitary CFT, wherein there exists a conformal frame such that the locations of a pair of operators in the correlator are related to the locations of the complementary pair by hermitian conjugation in radial quantization. In the Euclidean section, or when $z^* = \bar{z}$, these kinematics are given by the diagonal limit of $z = \bar{z}$. In the Lorentzian section, or when $z,\bar{z} \in(0,1)$, these kinematics are given by the ``self-dual line" of $z = 1-\bar{z}$. In this work, we will focus on OPE measures evaluated at the self-dual point of $z = \bar{z} = 1/2$.

We can rewrite the conformal block decomposition as an integral over a positive measure:
\ba
\mathcal{G}(z,\bar{z}) = \int_0^\infty d\Delta dJ_2\;\mu^\star(\Delta,J_2) \frac{G_{\Delta,\ell}(z,\bar{z})}{G_{\Delta,\ell}(z^\star,\bar{z}^\star)},
\ea
where $\ell = \left(\sqrt{J_2 + (\frac{d-2}{2})^2}-\frac{d-2}{2}\right) $ and 
\ba
\mu^\star(\Delta,J_2) = \delta(\Delta)\delta(J_2) + \sum_{\Delta'>0,\ell'} \delta(\Delta-\Delta') \delta(J_2 -J_2') a_{\Delta',\ell'}G_{\Delta',\ell'}(z^\star,\bar{z}^\star)
\ea
is a discrete measure weighted by squared and normalized OPE coefficients $a_{\cO}$ with the sum running over all primary operators $\cO \in \phi \times \phi$ labeled by $\Delta$ and $J_2$. This measure is positive by definition given our choice of $(z^\star,\bar{z}^\star)$ and the positivity of squared OPE coefficients in a unitary CFT. 

We aim to characterize this OPE measure by its moments, defined as
\ba
\nu_{m,n}(z^\star,\bar{z}^\star) = \int_0^\infty d\Delta dJ_2\;\Delta^m J_2^n \mu^\star(\Delta,J_2).
\ea
By convention, when $(z^\star,\bar{z}^\star) = (1/2,1/2)$, we will suppress the position dependence of the moments and the superscript $\star$ of the measure. We will also suppress the angular momentum moment index of $\nu_{m,n}$ when evaluating only scaling moments so that $\nu_m = \nu_{m,0}$. 

In appendix \ref{opemomentsproofs}, we prove that the double moment sequence $(\nu_{m,n})_{m,n\geq0}$ is Stieltjes determinant, and thus uniquely determines the underlying OPE measure $\mu(\Delta,J_2)$. Intuitively, this means that the bounds we derive on moments directly constrain CFT data as it arises in the OPE measure.

\subsection{Bounds from crossing and unitarity}

An associative OPE yields scalar four-point functions which are invariant under permutations of the external operators, expressed by equating
\ba
\mathcal{G}(u,v) = \left(\frac{u}{v}\right)^{\Df}\mathcal{G}(v,u) = \mathcal{G}(u/v,1/v),
\ea
with the OPE channels labeled s, t, and u respectively. This condition constitutes crossing symmetry, and subtracting the OPE decompositions of two of the channels gives rise to a consistency condition on CFT data expressed as a sum rule. The s-t crossing sum rule is
\ba
\sum_{\Delta,\ell} a_{\Delta,\ell} \left( 
 u^{-\Df}G_{\Delta,\ell}(u,v) - v^{-\Df}G_{\Delta,\ell}(v,u) \right) =\sum_{\Delta,\ell} a_{\Delta,\ell} F_{\Delta,\ell}(u,v)  = 0,
\ea
where we have multiplied through by $u^{-\Df}$ as a convention so that the crossing vector $F_{\Delta,\ell}(u,v)$ is antisymmetric under $u \leftrightarrow v$. Taylor expanding the crossing vector around $z = \bar{z} = 1/2$ gives a countable set of constraints order-by-order in the series\footnote{We denote evaluation at the self-dual point $z=\bar{z} =1/2$ by suppressing the dependence on the position variables.}
\ba
\sum_{m,n}  L_\mu\left[  \frac{ \partial^m_z \partial^n_{\bar{z}}F_{\Delta,\ell}}{G_{\Delta,\ell}}  \right] \frac{(z-1/2)^m(\bar{z}-1/2)^n}{m! n!} = 0,
\ea
 with all terms of even $n+m =\Lambda$ identically vanishing by the $u \leftrightarrow v$ antisymmetry of $F_{\Delta,\ell}$. 
 
 We can further decompose each Taylor coefficient into a sum over normalized derivatives of the conformal block:
\ba
L_\mu\left[  \frac{ \partial^m_z \partial^n_{\bar{z}}F_{\Delta,\ell}}{G_{\Delta,\ell}}  \right] = L_\mu\left[\sum_{ i j} c^{(m,n)}_{ij}(\Df) g_{i j} (\Delta,\ell)\right]=\sum_{ij}c^{(m,n)}_{ij}(\Df) g_{ij}=0,
\label{constraintcoeff}
\ea
where
\ba
g_{i j}(\Delta,\ell) = \frac{\partial_z^i \partial_{\bar{z}}^j G_{\Delta,\ell}}{ G_{\Delta,\ell}},
\label{pretaylorcoeff}
\ea
$ c^{(m,n)}_{ij}(\Df)$ are real coefficients depending only on the external scaling dimension $\Df$, and $L_\mu[g_{i j}(\Delta,\ell) ] = g_{ij}$ are Taylor coefficients of the correlation function expanded around $z = \bar{z} = 1/2$:
\ba
\mathcal{G}(z,\bar{z}) = \sum_{ij} \frac{g_{ij}}{i! j!} (z-1/2)^i (\bar{z}-1/2)^j.
\ea

Note that the $z \leftrightarrow\bar{z}$ symmetry of the conformal block implies $g_{ij} = g_{ji}$. Additionally, we find that $g_{ij}>0$ for all $i,j\geq0$. This can be seen by applying derivatives to the expansion of the conformal block in powers of $z,\bar{z}$ written in eq.~(78) in the appendix of~\cite{Fitzpatrick:2012yx}, noting that each term in the expansion is positive for all derivative orders. The 1d analog of these Taylor coefficients were previously studied in~\cite{Arkani_Hamed_2019}, where the authors used them to analyze the conformal bootstrap from the perspective of positive geometry. It is easy to use crossing symmetry to derive relations between these Taylor coefficients by evaluating the constraint coefficients in eq.~(\ref{constraintcoeff}). 

At a given order $\Lambda$ we can compute $\frac{\Lambda +1}{2}$ independent relations. For the first few orders we have:
\begin{align}
\underline{\Lambda = 1:} \qquad\qquad &0 = g_{10}-2 \Df g_{00}, \label{firstrel}\\
\underline{\Lambda = 3:} \qquad\qquad &0 = 16 \Delta _{\phi }^3g_{00}-16 \Delta _{\phi }g_{00}-6  \Delta _{\phi }g_{20}+g_{30}, \\
&0 = 16 \Delta _{\phi }^3g_{00}-4\Delta _{\phi }  g_{11} -2 \Delta _{\phi } g_{20}+g_{21}.
\end{align}
We would like to interpret these relations in terms of our classical moment variables by approximating
\ba
g_{ij} = \sum_{p,q} b_{p,q}^{(i,j)} \nu_{p,q} + L_\mu\left[O\left(\frac{1}{\Delta}\right)\right],
\ea
where $b_{p,q}^{(i,j)}$ are real coefficients and $L_\mu\left[O\left(\frac{1}{\Delta}\right)\right]$ is an error term arising from null state contributions to the conformal block, which we will bound in the following sections.

\subsubsection{Bound on average scaling dimension}

Let us warm up by deriving a bound on the first normalized moment $\langle\Delta\rangle = \nu_1/\nu_0$, which describes the average scaling dimension of the OPE measure $\mu(\Delta,J_2)$ at $z = \bar{z} =1/2$. To derive this bound, we need only the constraint of eq.~(\ref{firstrel}) and some numerical analysis of the conformal block. Write $G_{\Delta,\ell}(z,\bar{z}) = (4 r )^\Delta h_{\Delta,\ell}(z,\bar{z})$ where $r = \sqrt{\rho(z)\bar{\rho}(\bar{z})}$ is the radial coordinate and $h_{\Delta,\ell}(z,\bar{z})$ is the regulated conformal block. Now, compute
\ba
g_{10}(\Delta,\ell) &= \left. \frac{\partial_z G_{\Delta,\ell}(z,\bar{z})}{G_{\Delta,\ell}(z,\bar{z})} \right|_{z = \bar{z} =1/2}\\
&=\sqrt{2} \Delta + \frac{ \partial_z h_{\Delta,\ell} }{h_{\Delta,\ell}}.
\ea

The term involving the regulated conformal block is not known in closed form in general dimensions, and can only be computed up to a finite pole order by recursion relations for a given spin. Doing so numerically for a large number of spins and derivatives, we find the bound
\ba
0\leq \frac{\partial_z^m\partial_{\bar{z}}^n h_{\Delta,\ell}}{h_{\Delta,\ell}}\leq \frac{\partial_z^m\partial_{\bar{z}}^n h_\ell^\infty}{h_\ell^\infty}, 
\label{regbound}
\ea
which holds for all scaling dimensions satisfying $\Delta \geq \ell + d - 2$ as well as at the identity in spatial dimensions $d \leq 4$.\footnote{It can potentially be violated in $d \leq 4$ if there are exchanged scalars with $(d-2)/2 < \Delta < d - 2$ close to the unitarity bound.} 

More specifically, we find the value at the identity is $\frac{\partial_z^m\partial_{\bar{z}}^n h_{0,0}}{h_{0,0}} = 0$ for all $m+n >0$. Using the explicit form of $h^\infty_\ell(z,\bar{z})$ in eq.~(\ref{asymregblock}), we find
\ba
\sqrt{2} \Delta \leq g_{10}(\Delta,\ell) \leq \sqrt{2} \Delta + \left(\frac{3}{4}-\frac{1}{\sqrt{2}}\right) d
\label{intermediatebound}
\ea
for $\Delta\geq \ell + d-2$, and $g_{10}(0,0) = 0$. Using the positivity of the measure $\mu(\Delta,J_2)$, we can apply $L_\mu$ to eq.~(\ref{intermediatebound}) without affecting the inequality signs to find
\ba
\sqrt{2} \nu_1 \leq g_{10} \leq \sqrt{2} \nu_1 + \left(\frac{3}{4} - \frac{1}{\sqrt{2}}\right)d (\nu_0 -1),
\label{premomentbound}
\ea
where, for the second term of the RHS, we computed
\ba
L_\mu[ g_{10}(\Delta,\ell) - \sqrt{2}\Delta]&=L_{\{0\}}[ g_{10}(\Delta,\ell) - \sqrt{2}\Delta] + L_{\mu/\{0\}}[ g_{10}(\Delta,\ell) - \sqrt{2}\Delta ]\\
&\leq L_{\mu/\{0\}}[ 1 ]  \left(\frac{3}{4} - \frac{1}{\sqrt{2}}\right)d= \left(\frac{3}{4} - \frac{1}{\sqrt{2}}\right)d (\nu_0 -1)
\ea
with $\mu/\{0\}$ denoting the OPE measure with the identity operator subtracted and the measure $\{0\}$ is a normalized delta mass at $\Delta = J_2 = 0$. Using eq.~(\ref{firstrel}) and rearranging eq.~(\ref{premomentbound}), we find, for any correlator with a scalar gap $\Delta_\mathrm{gap} \geq d-2$ in $d \leq 4$, the following holds:
\ba
\boxed{\frac{d}{8}\left(4-3\sqrt{2}\right) \frac{( \nu_0 -1)}{\nu_0} +\sqrt{2} \Df \leq \frac{\nu_{1}}{\nu_0} \leq \sqrt{2} \Df}
\label{firstbound}
\ea

In dimensions $d>4$, we can still bound derivatives of the regulated block by some constant for $\Delta \geq \frac{d-2}{2}$, so more generally we have the asymptotic of $\nu_1/\nu_0 = \sqrt{2}\Df + O(1)$, or that the average scaling dimension of $\mu(\Delta,J_2)$ grows linearly with $\Df$ at a universal rate. The order at which this statement fails to be projective, in that it involves terms which are not of the form $\nu_k/\nu_0$, is subleading in the heavy limit of $\Df \to \infty$. This is seen by the fact that $0<\frac{\nu_0 -1}{\nu_0} <1$ for $\nu_0>1$, which is the case for any unitary CFT. Lastly, we remark that this result extends to moments the asymptotic constraints from ``reflection symmetry" previously observed in~\cite{Kim2015-sg, Paulos2016-mh}. 

\subsubsection{Leading-order bounds in the heavy limit}
\label{sec:lobounds}

We will now direct our attention to the higher scaling moments $\nu_n/\nu_0$ for $n>1$ in the heavy limit. Our aim is to derive the following bound:
\ba
\boxed{2^{n/2} \leq \frac{\nu_n}{\nu_0} \Df^{-n} + O\left( \frac{1}{\Df} \right) \leq 2^{3n/2 -1},}
\label{momentbound}
\ea
where the error term indicates that this bound may be violated by terms that decay as $1/\Df$ as $\Df \to \infty$. Before we dive into the derivation, a few preliminaries are in order:

\paragraph{Diagonal limit} To derive eq.~(\ref{momentbound}), we will make use of the crossing equation in the diagonal limit of $z = \bar{z}$. This choice will greatly simplify our derivation by removing the spin dependence of the asymptotic conformal block so that we are only working with relations between scaling moments. Since crossing symmetric correlators are also crossing symmetric in the diagonal limit, the bounds we derive with these relations are necessarily true. In the following section, we will further derive bounds on spinning moments to show that any off-diagonal constraint gives only new subleading corrections to moments in the heavy limit. In other words, eq.~(\ref{momentbound}) is sharp up to the given error terms. 

For notation, we will write the diagonal block as $G_{\Delta,\ell}(z) = G_{\Delta,\ell}(z,z)$ and the Taylor coefficient as $g_n = \left.L_\mu\left[ \frac{\partial_z^n G_{\Delta,\ell}(z)}{G_{\Delta,\ell}(z)}\right]\right|_{z=1/2}$. Expanding the diagonal crossing vector around $z = 1/2$ then gives the following constraints for odd $\Lambda$:
\ba
\sum_{n=0}^\Lambda \binom{\Lambda}{n}  \frac{2^{\Lambda -n} \Gamma \left(1-2 \Delta _{\phi }\right)}{\Gamma \left(n-\Lambda -2 \Delta _{\phi }+1\right)} g_{ n} = 0.
\label{diagconstraint}
\ea

\paragraph{Inverse moments} We would like to derive a simple bound on the $L_\mu[ O(\frac{1}{\Delta})]$ error terms that appear in our moment relations. This will allow us to safely ignore these terms in the heavy limit, further simplifying our analysis. To do this, consider the inverse moment defined as $\nu_{-1} = L_{\mu/\{0\}}[\Delta^{-1}]$, where we have subtracted the identity contribution from the measure to avoid any divergences at $\Delta = 0$. Writing $p_\Delta = a_\Delta G_\Delta(1/2)$, we explicitly have
\ba
\frac{\nu_{-1}}{\nu_{0}-1} = \frac{p_{\Delta_\mathrm{gap}} \Delta_\mathrm{gap}^{-1} + \sum_{\Delta>\Delta_\mathrm{gap} } p_\Delta \Delta^{-1} }{p_{\Delta_\mathrm{gap}}  + \sum_{\Delta>\Delta_\mathrm{gap} } p_\Delta},
\ea
where we have separated off the operator at the gap of the spectrum. Since $\Delta_\mathrm{gap}^{-1} > \Delta^{-1}$ for all $\Delta > \Delta_\mathrm{gap}$, we can bound
\ba
\frac{ \sum_{\Delta>\Delta_\mathrm{gap} } p_\Delta \Delta^{-1} }{\sum_{\Delta>\Delta_\mathrm{gap} } p_\Delta  } < \Delta_\mathrm{gap}^{-1}.
\label{gapbound}
\ea

Note that for any positive real numbers $a,b,c,d$ satisfying $\frac{a}{c} \geq \frac{b}{d}$, the following inequality holds:
\ba
\frac{a+b}{c+d} \leq \frac{a}{c}.
\label{simpleineq}
\ea
Letting $a =p_{\Delta_\mathrm{gap}} \Delta_\mathrm{gap}^{-1} $, $b = \sum_{\Delta>\Delta_\mathrm{gap} } p_\Delta \Delta^{-1} $, $c =p_{\Delta_\mathrm{gap}} $, $d =\sum_{\Delta>\Delta_\mathrm{gap} } p_\Delta  $, and using eq.~(\ref{gapbound}), we conclude
\ba
\frac{\nu_{-1}}{\nu_0-1} < \Delta_\mathrm{gap}^{-1},
\label{inversemomentbound}
\ea
%or that $L_\mu[O(\frac{1}{\Delta})]/\nu_0= O(\Delta_{\mathrm{gap}}^{-1})$.
or that $L_{\mu/\{0\}}[O(\frac{1}{\Delta})]/(\nu_0-1)= O(\Delta_{\mathrm{gap}}^{-1})$. Further, we use the results of this section to prove a stronger bound on inverse moments in appendix \ref{app:inversebound}.

\paragraph{Jensen's inequality} 
A very useful result in probability theory is Jensen's inequality, which states the following: Let $g(\bullet)$ be a convex function on an interval $I \in \mathbb{R}$, and $X$ a random variable taking values in $I$, then $E[g(X)] \geq g(E[X])$, where $E[X]$ denotes the normalized expectation value of $X$.  

If we let $I = [0,\infty)$ and view $\Delta$ as a random variable distributed according to the positive density $\mu(\Delta)$, then Jensen's inequality asserts that moments satisfy
\ba
\frac{\nu_n}{\nu_0} \geq \left(\frac{\nu_1}{\nu_0} \right)^n,
\label{jensensmoment}
\ea
for $n \geq 0$, since $g(X) =X^n$ is convex for $X \in I$. Eq.~(\ref{jensensmoment}) can also be derived from the more primitive bound arising from Hankel matrix positivity given by
\ba
\nu_{n} \nu_{n+2} \geq \nu_{n+1}^2
\label{hankelmomentbound}
\ea
for $n\geq0$, which follows directly by imposing Sylvester's criterion on the leading minor determinant of each shifted Hankel matrix $H^{(n)}_\infty \succeq 0$. When eq.~(\ref{hankelmomentbound}) is saturated for all $n\geq m$, the Hankel matrix $H^{(m)}_\infty$ becomes singular, indicating that the tail of the moment sequence $\{ \nu_n\}_{n\geq m}$ lies on the boundary of the convex moment cone carved out by total Hankel matrix positivity.

\paragraph{Bound on moment growth}
Before we can derive sharp bounds on normalized classical moments $\nu_n/\nu_0$, we need to bound their growth in the large $\Df$ limit. Namely, we want to show that
\ba
\frac{\nu_n}{\nu_0} = O(\Df^{n+\epsilon})
\label{momentgrowth}
\ea
for all $n>0$ and $\epsilon>0$. For the purposes of this paper, we will show a simplified argument in the case of $d = 1$, however higher dimensional generalizations follow from the work of~\cite{Mukhametzhanov:2018zja}. Consider the spectral density defined as
\ba
\varrho(\Delta) = \sum_{\Delta'}a_{\Delta'} C(\Delta')  \delta(\Delta - \Delta'),
\ea
where $C(\Delta) = 4^\Delta\sqrt{\frac{\Delta}{\pi}} $. In~\cite{Qiao:2017xif}, the authors proved the large $\Delta_0$ asymptotic of
\ba
Q(\Delta_0) = \int_0^{\Delta_0} d\Delta\;\varrho(\Delta) \sim \frac{\left(4 \Delta _{\phi }\right) \Delta _0^{4 \Delta _{\phi }}}{\Gamma \left(2 \Delta _{\phi }+1\right){}^2},
\label{tauberianthmasym}
\ea
which is valid for $\Delta_0 \gg \Df$.
We will use this asymptotic to prove eq.~(\ref{momentgrowth}). 

Define $p(\Delta) = \varrho(\Delta)C(\Delta)^{-1}G_\Delta(1/2)$ and write
\ba
\nu_n^{<} = \int_{0}^{\Df^{1 + \epsilon/n}} d\Delta p(\Delta) \Delta^n, \quad \quad \nu_n^{>} = \int_{\Df^{1 + \epsilon/n}}^\infty d\Delta p(\Delta) \Delta^n,
\ea
with some $\epsilon > 0$, noting $\frac{\nu_n}{\nu_0} = \frac{\nu_n^< + \nu_n^>}{\nu_0^< + \nu_0^>}$. It is easy to see that
\ba
\frac{\nu^<_n}{\nu^<_0} = \frac{\int_{0}^{\Df^{1 + \epsilon/n}} d\Delta p(\Delta) \Delta^n}{\int_{0}^{\Df^{1 + \epsilon/n}} d\Delta p(\Delta) } \leq \Df^{n+\epsilon}
\ea
by bounding $\Delta^n \leq  \Df^{n+\epsilon}$ for $\Delta \in [0,\Df^{1+ n/\epsilon}]$. For the tail integral, we note $\varrho(\Delta) = Q'(\Delta)$ and write 
\ba
\nu^>_n = \int_{ \Df^{1 + n/\epsilon} }^\infty d\Delta Q'(\Delta) C(\Delta)^{-1} G_{\Delta}(1/2) \Delta^n.
\ea
We can now use integration by parts and the asymptotics of eq.~(\ref{tauberianthmasym}) and eq.~(\ref{confblockasym}) to evaluate this integral explicitly. The use of the asymptotic is justified here as all scaling dimensions in the integrand are bounded from below by $\Df^{1+ \epsilon/n} \gg \Df$ for $\Df$ sufficiently large. 

Expanding the result around $\Df \to \infty$ then yields
\ba
\frac{\nu_n^{>}}{\nu_0^{>}} = \Df^{n+\epsilon}\left( 1 + O\left(\frac{1}{\Df} \right) \right).
\ea
Clearly, if both $\frac{\nu^<_n}{\nu^<_0} = O(\Df^{n+\epsilon})$ and $\frac{\nu^>_n}{\nu^>_0} = O(\Df^{n+\epsilon})$, then $\frac{\nu_n}{\nu_0} = O(\Df^{n+\epsilon})$ as we can choose whichever average is largest and bound the total using eq.~(\ref{simpleineq}). Thus, we conclude the proof of eq.~(\ref{momentgrowth}).

\paragraph{Proof of the asymptotic moment bounds}
Now, let us derive the bounds in eq.~(\ref{momentbound}).

\underline{Lower bound:} We can use eq.~(\ref{jensensmoment}) with $\nu_1/ \nu_0 = \sqrt{2}\Df + O(1)$ to compute
 \ba
\frac{\nu_n}{\nu_0} \geq (\sqrt{2} \Df )^n + O(\Df^{n-1} ).
 \ea
Dividing both sides by $\Df^n$ gives the lower bound in eq.~(\ref{momentbound}). As a corollary, using the bound on growth rate in eq.~(\ref{momentgrowth}) and the lower bound here, we can fix some $1>\epsilon > 0$ so that $\nu_{p}$ grows faster than $\nu_{q}$ for all $p>q$ as $\Df \to \infty$. In turn, we can compute an asymptotic expansion of $g_n$ in the large $\Df$ limit order-by-order in $\nu_k$.

\underline{Upper bound:} We want to understand the diagonal constraints of eq.~(\ref{diagconstraint}) in the limit of large $\Df$. The $\Df$-dependent coefficients can be easily expanded to find
\ba
\frac{2^{\Lambda -n} \Gamma \left(1-2 \Delta _{\phi }\right)}{\Gamma \left(n-\Lambda -2 \Delta _{\phi }+1\right)} = (-4 \Df)^{\Lambda - n}\left(1 + O\left(\frac{1}{\Df} \right) \right).
\label{leadingcoeff}
\ea

To compute an asymptotic for Taylor coefficients, we write $G_{\Delta,\ell}(z) = (4 \rho(z))^\Delta h_{\Delta,\ell}(z)$ with 
\ba
h_{\Delta,\ell}(z) = (1-\rho(z)^2)^{-\frac{d}{2}} + O\left( \frac{1}{\Delta} \right).
\ea
It is then easy to compute
\ba
 \left.\frac{\partial_z^n G_{\Delta,\ell}(z)}{G_{\Delta,\ell}(z)} \right|_{z = 1/2} &=\left. \sum_{k}^n \binom{n}{k} \frac{\partial_z^k \rho(z)^\Delta}{\rho(1/2)^{\Delta}} \frac{\partial_z^{n-k}h_{\Delta}(z)}{h_{\Delta}(1/2)}  
 \right|_{z = 1/2}\\
&= \left(2^{3/2}\Delta\right)^n + O(\Delta^{n-1} ).
\ea
Thus, we have the leading piece of
\ba
g_n = 2^{3n/2} \nu_n + O(\nu_{n-1} ).
\label{leadingtaylor}
\ea
Plugging eq.~(\ref{leadingcoeff}) and eq.~(\ref{leadingtaylor}) into eq.~(\ref{diagconstraint}) then yields
\ba
\sum_{n=0}^\Lambda \binom{\Lambda}{n} \left( - \sqrt{2}\Df\right)^{\Lambda - n} \nu_n = O\left(\nu_{\Lambda -1}\right),
\label{leadingasymptoticconstraint}
\ea
or more intuitively
\ba
 L_\mu[ ( \Delta - \sqrt{2}\Df)^\Lambda  + O(\Delta^{\Lambda - 1} )] = 0,
\ea
which is the constraint that leading odd central moments vanish as $\Df \to \infty$.

Now that we have established the asymptotic constraints from crossing, we can turn to constraints from unitarity. Intuitively, one expects bounds on moment sequences to coincide with the boundary of the convex moment cone arising from Hankel matrix positivity. This is indeed the case, as we will show by explicitly constructing ``extremal" moment sequences which lie on the moment cone and verifying that they satisfy crossing symmetry. 

Via Sylvester's criterion, demanding that the determinants of all the leading minors of $H^{(0)}_\infty$ or $H^{(1)}_\infty$ vanish gives rise to the extremal moment sequences of 
\ba
\left(\frac{\nu_n}{\nu_0} \right)_{(0)} = \left( \frac{\nu_1}{\nu_0} \right)^n \quad \mathrm{or} \quad\left( \frac{\nu_n}{\nu_0} \right)_{(1)} = \frac{\nu_2}{\nu_0}\left( \frac{\nu_2}{\nu_0} \right)^{n-2},
\label{skeletonsequences}
\ea
respectively. Note that the $(0)$ moment sequence is just the lower bound in eq.~(\ref{momentbound}). The $(1)$ sequence is more non-trivial, and depends on an unfixed $\nu_2/\nu_0$ moment. We can compute extremal solutions of $\nu_2/\nu_0$ by asymptotically solving a system of equations given by eq.~(\ref{leadingasymptoticconstraint}) up to order $\Lambda =3$ and eq.~(\ref{skeletonsequences}) at $n = 3$. In total, these constraints give rise to the quadratic equation
\ba
\frac{1}{\Df^4 } \left(8  \Delta _{\phi }^4-6  \Delta _{\phi }^2\left(\frac{\nu _2 }{\nu_0}\right)+\left(\frac{\nu _2 }{\nu_0} \right)^2\right) = O\left( \frac{1}{\Df} \right),
\ea
with solutions
\ba
\left(\frac{\nu_2}{\nu_0}\right)_- = 2 \Df^2 + O (\Df) \quad \mathrm{and} \quad  \left(\frac{\nu_2}{\nu_0} \right)_+ = 4 \Df^2 + O (\Df).
\ea

We note that $\left(\nu_2/\nu_0\right)_-$ already saturates our lower bound, and $\left(\nu_2/\nu_0\right)_+$ is the largest value of $\nu_2/\nu_0$ (up to error terms) which satisfies the constraints from crossing symmetry and unitarity. Plugging $\left(\nu_2/\nu_0\right)_+$ into our equation for $\left(\nu_n/\nu_0\right)_{(1)}$ extends the extremal sequence to
\ba
\left(\frac{\nu_n}{\nu_0}\right)_+ = 2^{3n/2} \Df^n + O(\Df^{n-1}),
\ea
which we call the maximal sequence, and we call
\ba
\left(\frac{\nu_n}{\nu_0}\right)_- = 2^{n/2} \Df^n + O(\Df^{n-1}),
\ea
the minimal sequence. Moreover, we can plug these extremal sequences into eq.~(\ref{leadingasymptoticconstraint}) to verify they are indeed crossing symmetric at all orders of $\Lambda$. Since we have constructed these sequences to lie on the boundary of the moment cone carved out by unitarity, as well as extremize their values subject to crossing symmetry, it follows that any normalized $n$-th moment satisfies
\ba
\left(\frac{\nu_n}{\nu_0} \right)_-\leq \frac{\nu_n}{\nu_0} \leq \left(\frac{\nu_n}{\nu_0}\right)_+
\ea
for all $n \geq 0$. Dividing by $\Df^n$ gives eq.~(\ref{momentbound}). This concludes the proof.

So far we considered the leading behavior of the moments in the heavy limit $\Df \rightarrow \infty$. We can also use similar methods to obtain constraints on subleading terms in the large $\Df$ expansion. We describe some of these constraints in appendix~\ref{app:subleading}.

\subsubsection{Spin moments and covariance bound}
\label{sec:spinmoments}

In this subsection, we will derive the following bound in the heavy limit:
\ba
\boxed{0 \leq \frac{\mathrm{Cov}(\Delta,J_2)}{\Df^2} + O\left(\frac{1}{\Df} \right) \leq \frac{d-1}{\sqrt{2}}, }
\label{covbound}
\ea
where $\mathrm{Cov}(\Delta,J_2) = \frac{L_\mu[\Delta J_2]}{L_\mu[1]} - \frac{L_\mu[\Delta]}{L_\mu[1]}\frac{L_\mu[J_2]}{L_\mu[1]}$ is the \textit{covariance} of two random variables $\Delta,J_2 \in [0,\infty)$ distributed with bi-variate density $\mu(\Delta,J_2)$. Once again, we begin with a few preliminary results:

\paragraph{Bound on growth of spinning and mixed moments} In the previous section, we derived upper bounds on scaling moments in the heavy limit concluding that $\frac{\nu_n}{\nu_0} = a_n \Df^n + O\left(\Df^{n-1} \right)$ for all $n \geq 0$ with some real $2^{n/2}\leq a_n \leq  2^{3n/2 -1}$. We would like to extend this result to spinning and mixed moments so that we can generally write
\ba
\frac{\nu_{mn}}{\nu_0} = a_{mn} \Df^{m+n} + O\left(\Df^{m + n -1}\right).
\label{loform}
\ea
This fact is non-trivial, as if we na\"ively bound $J_2 \leq \Delta^2$, then we only have $\frac{\nu_{mn}}{\nu_0} = O\left(\Df^{m + 2n} \right)$, which is notably weaker than eq.~(\ref{loform}). 

To derive the stronger bound, first note that $g_n/\nu_0 = O(\nu_n/\nu_0) = O(\Df^n)$ (from eq.~(\ref{leadingtaylor})). We can use the chain rule to expand diagonal Taylor coefficients in terms of non-diagonal Taylor coefficients as
\ba
g_n = \sum_{k}^n \binom{n}{k} g_{k,n-k}.
\ea
Since each term in this sum is manifestly positive, no cancellations can occur between the Taylor coefficients, thus $g_n/\nu_0 = O(\Df^n)$ implies that $g_{k,n-k}/\nu_0 = O(\Df^n)$ for all $n \geq 0$ and $k \in [0,n]$. Writing $n = i +j$ and $k = i$ then yields $g_{ij}/\nu_0 = O\left(\Df^{i+j}\right)$ for all $i,j \geq 0$. Now, note that $\Omega_-^2$ contains a second-order piece of the form $\sim(1 -\eta^2)\partial_\eta^2$ which vanishes after taking the diagonal limit (which sets $\eta = 1$), so $\Omega_-^2$ acts as a differential operator of order 1 after taking $z = \bar{z} =1/2$. 

Using the crossing equation, we can write
\ba
\frac{\nu_{mn}}{\nu_0} &=\frac{1}{\nu_0} \left.\left( \Omega_+ +\frac{d}{2} \right)^m \left( \Omega_-^2 - \left(\frac{d-2}{2}\right)^2 \right)^n \left[ \left(\frac{z \bar{z}}{(1-z)(1-\bar{z})} \right)^{\Df} \mathcal{G}(1-z,1-\bar{z})  \right]\right|_{z = \bar{z} =1/2}\\
&= \frac{1}{\nu_0}\sum_{j,k}^{m+n} c_{j,k}^{(m,n)} \Df^{m+n-j-k}g_{j,k}+ O\left(\Df^{ m + n -1}\right)  + O\left(\frac{1}{\Delta_\mathrm{gap} } \right) \\&= a_{mn} \Df^{m+n} + O\left(\Df^{m+n-1} \right)
\ea
for coefficients $c_{j,k}^{(m,n)}, a_{mn} \in \mathbb{R}$, where, in the second line, we  used the estimate (see eq.~(\ref{inversemomentbound})) of $L_\mu\left[ O\left(\frac{1}{\Delta} \right) \right] = O\left(\frac{1}{\Delta_{\mathrm{gap}} } \right)$ and the observation that terms which grow faster than $O\left( \Df^{m+n}\right)$ vanish after evaluating $z = \bar{z} = 1/2$. 

Thus, we conclude the proof of eq.~(\ref{loform}). We also remark that, away from the diagonal limit, the second-order piece of $\Omega_-^2$ survives, so we have $\frac{\nu_{mn}(z \neq \bar{z})}{\nu_0(z \neq \bar{z})} = O\left(\Df^{m + 2n} \right)$ as our na\"ive bound suggested. 

\paragraph{Cylinder coordinates}
Since diagonal constraints have no dependence on spin moments, it is useful to adopt a coordinate system wherein we can separate diagonal from off-diagonal constraints. One choice for this is cylinder coordinates (see eq.~(\ref{cylindercoords})), where taking derivatives of the crossing equation with respect to $\tau$ gives diagonal constraints, and derivatives with respect to $\theta$ gives off-diagonal constraints. 

Consider the crossing equation in cylinder coordinates, with the identity subtracted, as
\ba
\left(\frac{4}{(\cos (\theta )+\cosh (\tau ))^2}\right)^{\Delta _{\phi }}-\left(1-\frac{2 \cos (\theta )}{\cos (\theta )+\cosh (\tau )}\right)^{2 \Delta _{\phi }} = L_{\mu / \{0\}}\left[ \frac{F_{\Delta,\ell}(\tau,\theta)}{G_{\Delta,\ell}(\tau,\theta)}\right],
\ea
where we write $\frac{F_{\Delta,\ell}(\tau,\theta)}{G_{\Delta,\ell}(\tau,\theta)}$ using the asymptotic conformal blocks at large $\Delta$. Expanding both sides around $\theta = 0$ and evaluating at $\tau = \log(3-2\sqrt{2})$ gives the following asymptotic constraints at order $\theta^2$ and $\theta^4$ respectively:
\ba
0&=\left(2 \sqrt{2}-3\right) d \left(\nu _0-1\right)+8 \nu _0 \Delta _{\phi }-4 \sqrt{2} \nu _{10}
\ea
and
\ba
0&= \left(\nu _0-1\right) \left(\left(36 \sqrt{2}-51\right) d^2+d \left(24 \left(2 \sqrt{2}-3\right) \Delta _{\phi }-40 \sqrt{2}+54\right)+16 \left(24 \Delta _{\phi }^2-13 \Delta _{\phi }+3\right)\right)\\&-8 \nu _{10} \left(3 \left(3 \sqrt{2}-4\right) d+2 \sqrt{2} \left(6 \Delta _{\phi }-5\right)\right)-\frac{48 \nu _{01} \left(\left(2 \sqrt{2}-3\right) d+8 \Delta _{\phi }+4\right)}{d-1}+\frac{192 \sqrt{2} \nu _{11}}{d-1}\\&-32 \Delta _{\phi } \left(5-12 \Delta _{\phi }\right)-96 \nu _{20}.
\label{spinconstraint}
\ea
The order $\theta^2$ constraint has no dependence on spin, and is identical to those derived from the diagonal limit, so the first non-trivial constraint on spin moments from crossing is given by eq.~(\ref{spinconstraint}).

\paragraph{Proof of the covariance bound (\ref{covbound})}

Using our spin constraint from crossing and the leading-order behavior of moments, we can derive a bound on the leading term in the covariance of $\Delta$ and $J_2$. First, let $\frac{\nu_{mn}}{\nu_0} = a_{mn}  \Df^{m+n} + O(\Df^{m+n-1})$. Plugging this into eq.~(\ref{spinconstraint}) and taking the heavy limit gives
\ba
\left(\sqrt{2} a_{10}+a_{20}-4\right) (d-1)+4 a_{01}-2 \sqrt{2} a_{11}=O(\Df^{-1}).
\ea
From our diagonal constraints and bounds from unitarity, we know that $a_{10} = \sqrt{2}$ and $2\leq a_{20}\leq4$, implying the bound on spinning/mixed moments 
\ba
0\leq 2 \sqrt{2} a_{11} - 4 a_{01} +O( \Df^{-1})\leq 2(d-1).
\ea
We can rephrase this as a bound on the covariance $\frac{\text{Cov}( \Delta, J_2)}{\Df^2}= a_{11} - \sqrt{2}a_{01} + O( \Df^{-1})$, which gives eq.~(\ref{covbound}).

\paragraph{Remark on sharpness of (\ref{momentbound})} To conclude that eq.~(\ref{momentbound}) is sharp, we need to confirm that off-diagonal constraints do not additionally constrain the leading order behavior of scaling moments. In other words, we want to show that any constraints involving derivatives acting on the regulated conformal block are sub-leading. Indeed, we can explicitly check that $\partial_z^n \partial_{\bar{z}}^m h_\ell^\infty(z,\bar{z})|_{z=\bar{z} = 1/2} = P^{\lfloor\frac{n+m}{2}\rfloor}(J_2)$, where $P^{\lfloor\frac{n+m}{2}\rfloor}(J_2)$ is an $\lfloor\frac{n+m}{2}\rfloor$ degree polynomial in $J_2$. Since $J_2$ moments grow with $\Df$ at the same rate as $\Delta$ moments, and $\lfloor\frac{n+m}{2}\rfloor < n+m$, any constraints which arise from taking derivatives of the regulated conformal block are subleading in the heavy limit. Since the remaining part of the block depends only on variables which are symmetric under $z \leftrightarrow \bar{z}$, we can freely restrict to the diagonal limit without losing any leading order constraints. 

If we were studying moments not evaluated in the diagonal limit, then $J_2$ moments grow at double the rate of $\Delta$ moments. This means that scaling moments and spinning moments contribute to constraints on crossing at the same order in $\Df$, and we can no longer study them separately as we do in this paper. This is exactly what we expect in the double light cone limit, where crossing and unitarity strongly constrain the OPE distribution over twist $\tau = \Delta -\ell$.

\section{Saddles and deformations}
\label{sec:saddles}

In this section, we will focus on OPE distributions over scaling dimension, defining
\ba
\mu(\Delta)  \equiv  \sum_{J_2 } \mu(\Delta,J_2).
\ea
The moments of this distribution can be computed from a moment-generating function $M_{\Delta}(t)$ as $\nu_n = \partial_t^n M_{\Delta}(t)|_{t=0}$.

Let us first consider the formal limit of $\Df \to \infty$, where it will be convenient to work with the rescaled variable $\tilde{\Delta} = \Delta/\Df$.
The moments of $\tilde{\Delta}$ are then sharply bounded as in eq.~(\ref{momentbound}) as $2^{n/2} \leq \tilde{\nu}_n/\nu_0 \leq 2^{3n/2-1}$ for all $n$, where $\tilde{\nu}_n = \nu_n/\Df^n$.

The upper bound is saturated by the moment-generating function
\ba
M^{(+)}_{\tilde{\Delta}}(t) = \frac{1}{2}\left(1+ e^{2\sqrt{2} t}\right).
\ea
Similarly, the lower bound is saturated by the moment-generating function
\ba
M^{(-)}_{\tilde{\Delta}}(t) = e^{\sqrt{2} t}.
\ea
These ``extremal" moment-generating functions correspond to the asymptotic measures
\ba
\mu^{(+)}(\tilde{\Delta})= \frac{1}{2}\left( \delta(\tilde{\Delta}) + \delta(\tilde{\Delta}-2\sqrt{2})\right)
\ea
and
\ba
\mu^{(-)}(\tilde{\Delta}) = \delta(\tilde{\Delta}-\sqrt{2}).
\ea
While these measures are clearly unphysical, in that they would not give rise to an exactly crossing-symmetric OPE at finite $\Df$, the locations of the $\delta$-distributions should be viewed as describing the approximate weights and locations of the dominant operator contributions to the OPEs of extremely heavy correlators.

In general, an asymptotic moment-generating function will sit between these, with
\ba
M^{(-)}_{\tilde{\Delta}}(t) < M_{\tilde{\Delta}}(t) < M^{(+)}_{\tilde{\Delta}}(t).
\ea
Considering the structure as a sum of exponentials, we can write down a ``heavy" ansatz as
\ba
M_{\tilde{\Delta}}(t)  \sim \frac{1}{\nu_0 } \sum_K  h_K  e^{\alpha_K t + O(t^2)},
\ea
with positive weights $h_K$. Here we interpret $\alpha_K$ as the locations of ``saddle points" associated with sharp peaks in the OPE distribution at dimensions scaling linearly with $\Df$. As implied by our upper bound, we expect that $\mathrm{max}[ \{ \alpha_K \} ] = 2\sqrt{2}$ is our heaviest saddle and we take $\alpha_0 = 0$ as the saddle associated with the s-channel identity contribution. The $O(t^2)$ term in the exponential represents subleading corrections to this saddle point approximation that broaden and skew each $\delta$-distribution while maintaining its average at $\alpha_K$. 

We would like to further specify the form of this saddle decomposition for correlators with large identical external scaling dimension prepared in different theories. To this end, we would like to first discuss the OPE saddle structures that arise for correlators in a generalized free theory. We will see that OPE saddles are in 1-to-1 correspondence with higher-spin (HS) conformal blocks that decompose free correlators, representing families of multi-twist operators involving a fixed number of elementary fields.

\subsection{Generalized free fields and higher-spin conformal blocks}
\label{sec:gffsaddles}
A generalized free field (GFF) theory provides an important playground for studying the structure of heavy correlators. Our analysis considers different ways one can construct a heavy operator in the theory. The AdS bulk action of a GFF is given by a massive free scalar field 
\ba
S^{GFF} = \int_{AdS} \sqrt{g}\left( \frac{1}{2}(\partial\phi)^2 - \frac{1}{2}m^2 \phi^2 \right).
\label{bulkGFFaction}
\ea

In this theory, all correlators in the boundary CFT can be computed with Wick contractions, which correspond to the disconnected exchange of the field(s) through geodesic paths in the AdS bulk. Additionally, we can define normal ordered products of fields $\phi^N \equiv :\hat{\phi}^N:$ by taking their OPE and subtracting off singular terms. The scaling dimension of the product field is given by $\Delta_{\phi^N} = N \Df$. Since we can vary both $\Df$ and take an arbitrary number of normal ordered products, we can construct operators with identical scaling dimensions but different OPEs.

Consider a four-point correlator of the form
\ba
\mathcal{G}_N(z,\bar{z}) = \frac{\langle \phi^N(x_1) \phi^N(x_2) \phi^N (x_3)\phi^N(x_4)\rangle}{\langle\phi^N(x_1)\phi^N(x_2)\rangle \langle \phi^N(x_3) \phi^N(x_4)\rangle},
\label{fourpointgn}
\ea
where we have normalized by the product of two-point correlators
\ba
\langle\phi^N(x_1)\phi^N(x_2)\rangle &= \frac{N!}{x^{2\Df N}_{12}}.
\ea
The full correlator can be computed directly with Wick contractions corresponding to the propagation of fields along geodesic paths in AdS space between points $12 \to 34$, $13\to24$, and $14\to 23$. The result is  
\ba
\mathcal{G}_N(z,\bar{z}) = \sum^N_{K=0}\binom{N}{K}^2 \mathcal{H}_K\left( u^{\Df},\left(u/v\right)^{\Df}\right),
\ea
where
\ba
\mathcal{H}_n(x,y) &= \sum_{k=0}^n \binom{n}{k}^2 x^{n-k} y^k = x^n \, _2F_1\left(-n,-n;1;\frac{y}{x}\right)
\ea
are the so-called ``higher-spin" (HS) conformal blocks introduced in~\cite{Alday2016-kz}. 

Focusing our analysis on the OPE distribution over scaling dimension, we will restrict to the 1d kinematics of the diagonal limit $z = \bar{z}$ and define $P \equiv z^{2\Df}$ and $Q \equiv (1-z)^{2\Df}$. In these variables, the 1d correlator reads
\ba
\mathcal{G}_N(z) = \sum_{K=0}^{N} \binom{N}{K}^2 \mathcal{H}_K\left(P,P/Q  \right).
\label{free1dcorrelator}
\ea

To compute the decomposition of $\mathcal{G}_N$ in terms of 1d conformal blocks, we will make use of the $\alpha$-space identity from~\cite{Hogervorst:2017sfd}:
\ba
z^p(1-z)^{-q} = \sum_{j=0}^\infty \frac{(p)_j^2}{j!(2p-1+j)_j} {}_3 F_2\left(\genfrac{}{}{0pt}{0}{-j,\;2p-1+j,\;p-q}{p,\;p};1\right) k_{p+j}(z).
\label{alphaid}
\ea

Let us warm up with the $N=1$ case, or the GFF correlator of the elementary primary field $\phi$:
\ba
\mathcal{G}_1(z) = 1+P+\frac{P}{Q}.
\ea
Using the above identity, one can verify the decomposition into 1d blocks:
\ba
\mathcal{G}_1(z)= 1+\sum_n a^{GFF}_{n}[\Df] G_{2\Df+2n}(z)
\ea
with 
\ba
a^{GFF}_{n}[\Df] = \frac{2 \Gamma^2 \left(2 n+2 \Delta _{\phi }\right) \Gamma \left(2 n+4 \Delta _{\phi }-1\right)}{\Gamma (2 n+1) \Gamma^2 \left(2 \Delta _{\phi }\right) \Gamma \left(2 \left(2 n+2 \Delta _{\phi }\right)-1\right)},
\ea
where we recognize the only contributing operators as the double-twist family $[\phi \phi]_n$ with scaling dimensions $\Delta_n = 2\Df + 2n$ and OPE coefficients 
\ba
\lambda^2_{\phi \phi [\phi \phi]_n}  = a^{GFF}_{n}[\Df].
\ea

Moving on to general $N$, we note that each term in the sum over higher-spin conformal blocks is characterized by an overall power of $P^K$, giving rise to a 1d conformal block decomposition with a gap at $\Delta = 2\Df K$. Applying eq.~(\ref{alphaid}) to all the terms at each $K$ unveils a highly degenerate operator spectrum with $\Delta_n = 2\Df K +2 n$ for positive integer $n$. Thus, the OPE is of the form
\ba
\mathcal{G}_N(z) = 1+\sum_{K=1}^N \binom{N}{K}^2 \sum^\infty_{n=0} a_{K,n} G_{2\Df K + 2n}(z).
\ea

For a given $K$, the $a_{K,n}$ coefficients admit a closed form in terms of hypergeometric functions obtained from the expansion coefficients of eq.~(\ref{alphaid}): 
\ba
a_{K,n} = \sum^K_{m=0}\binom{K}{m}^2 \frac{(2\Df K)_{2n}^2}{(2n)!(4\Df K-1+2n)_{2n}} {}_3 F_2\left(\genfrac{}{}{0pt}{0}{-2n,\;4\Df K-1+2n,\;2\Df (K-m)}{2\Df K,\;2\Df K};1\right).
\ea
In the heavy limit of $\Df \to \infty$, $a_{K,n}$ becomes peaked around $n \sim \frac{2}{3} \Df K$ and is very well approximated as $a_{K,n} \sim a_n^{GFF}[\Df K]$. We can re-sum these dominant contributions to obtain the asymptotic correlator of
\ba
\mathcal{G}_N(z) \sim 1 + \sum_{K=1}^N \binom{N}{K}^2 \left( \left( \frac{P}{Q}\right)^K +P^K\right) = \mathcal{H}_N(1,P) + \mathcal{H}_N(1,P/Q)-1,
\ea
which agrees with the result from taking eq.~(\ref{free1dcorrelator}) in the approximation $\mathcal{H}_K\left(P,P/Q  \right) \approx (P/Q)^K + P^K$, i.e.~keeping the term $(P/Q)^K$ with the largest power of $1/Q$ along with its image under $x_1 \leftrightarrow x_2$ exchange in eq.~(\ref{fourpointgn}). Note that if we restrict to kinematics in the Euclidean section with $0<z<1$, then the powers of $P^K$ fall off exponentially in the heavy limit, and we find
\ba
\mathcal{G}_N(z) \sim \mathcal{H}_N(1,P/Q) = Q^{-N}\mathcal{H}_N(Q,P).
\ea
This asymptotic correlator has a spectrum with scaling dimensions $2\Df K +j$ for $K \in [1,N]$ and $j \in \mathbb{Z}^+$, rather than the standard double integer spaced spectrum we observed in the full correlator. This is because terms which are subleading in the heavy limit serve to subtract off the ``odd spin" operators that arise in the leading-order result. 

Based on the form of the full 1d correlator given in (\ref{free1dcorrelator}), we expect our classical moment-generating function to be of the form
\ba
M_{\tilde{\Delta}}(t) = \frac{1}{\mathcal{G}_N(z)}\left( 1 + \sum^N_{K=1} \binom{N}{K}^2 \mathfrak{M}_K(t,z)  \right),
\ea
where $\tilde{\Delta} = \Delta/(N\Df)$ and $ \mathfrak{M}_K(t,z)= e^{(N\Df)^{-1}(\Omega_+ +1/2) t} \mathcal{H}_K(P,P/Q)$ are the un-normalized classical moment-generating functions of an individual higher-spin conformal block. 

In the heavy limit, these leading moments read 
\ba
(N\Df)^{-j}\left( \Omega_+ +\frac{1}{2}\right)^j \mathcal{H}_K(P,P/Q) = \left( \left(\frac{2 (K/N)}{\sqrt{1-z}}\right)^j + O((N\Df)^{-1}) \right) \mathcal{H}_K(P,P/Q).
\ea
Resumming these leading terms into the classical moment-generating function gives
\ba
\mathfrak{M}_K(t,z)= e^{\frac{2 (K/N)}{\sqrt{1-z}} t + O(t^2)} \mathcal{H}_K(P,P/Q),
\ea
or equivalently that each higher-spin conformal block is associated with a single saddle located at $\Delta = \frac{2\Df K}{\sqrt{1-z}} + O(1)$. The locations of these saddles coincide with dominant operator contributions at $ n \sim \left( \frac{1}{\sqrt{1-z}} -1 \right) \Df K$.\footnote{For $z = 16/25$, the asymptotic conformal block is a constant and the locations of OPE saddles are just the peaks of the bare OPE coefficients at $n \sim \frac{2}{3}\Df K$.} This can also be made apparent if one directly studies the $a_{K,n}$ OPE coefficients multiplied by conformal blocks of dimension $2\Df K+2n$. However, the result obtained here required no knowledge of the exact CFT data, and instead emerged only from a leading-order moment analysis of the higher-spin conformal blocks. 

With this leading moment-generating function for each saddle known, let us consider the moments of the total measure generated by $M_{\tilde{\Delta}}(t)$. To make contact with our previous bootstrap results from crossing, we will restrict our analysis here to the self-dual point $z = 1/2$. If we take $\Df K \to \infty$ at $z=1/2$, the higher-spin conformal block $\mathcal{H}_K(P,P/Q) \sim 1$ and the saddles are located at $\De = 2\sqrt{2}\Df K$ for $K \in [1,N]$. Additionally, the value of the correlator goes as 
\ba
\mathcal{G}_N(1/2) \sim \mathcal{H}_N(1,1) = \binom{2N}{N}.
\ea

First, we consider the case $N = 1$ in the heavy limit. We find the total normalized $k-$th moments are given by $\nu_k/\nu_0 = \frac{1}{2}(2\sqrt{2} \Df)^k + O(\Df^{k-1})$, which match the leading upper bounds $\nu_k^{(+)}/\nu_0$. On the other hand, if we fix $\Df$ and take the long limit of $N\to\infty$, we find $\nu_k/\nu_0 = (\sqrt{2} N\Df)^k + O( N^{k-1})$, which match the moments $\nu_k^{(-)}/\nu_0$ saturating the leading lower bound we derived. This gives meaning to the extremal moment sequences $\nu_k^{(+)}/\nu_0$ and $\nu_k^{(-)}/\nu_0$ as the $N=1, \Df \to \infty$ and $N\to \infty$ limits of the GFF correlator, respectively. The latter case is more universal in that these moments are recovered for all $\Df$ in the long limit, while the $\nu_k^{(+)}/\nu_0$ sequence is only recovered when $N=1$. 

We note that the lack of a saddle associated to the identity in $\mu^{(-)}(\tilde{\Delta})$ is a result of non-identity operators dominating the correlator in the long limit. This dominance is apparent as $\mathcal{G}_N(1/2) \sim \binom{2N}{N} \gg 1$ as $N \to \infty$. Moreover, the binomial coefficient $\binom{N}{K}^2$ becomes sharply peaked around $K = N/2$ as $N \to \infty$, so saddles distributed around $K \sim N/2$ give the dominant contributions out of the infinite sum over saddles which arises in the long limit. We'll discuss the properties of this collective distribution over saddles in the long limit further below.

We can refine our picture of these saddles by estimating the $O(t^2)$ terms in the exponential of our moment-generating function. A simple way to do this is by making a smooth ansatz for the derived measure $\mu_K(\tilde{\Delta})$ obtained by taking the inverse Laplace transform of $\mathfrak{M}_K(t,z)$, and matching the moments which parametrize it. As we will see in more detail below, the best ansatz for saddles in the heavy limit is given by a Gaussian
\ba
\mu_K(\tilde{\Delta}) \sim  \frac{\mathcal{H}_K}{\sigma_K\sqrt{2\pi}}\exp\left({\frac{-(\tilde{\Delta}-m_K)^2}{2\sigma_K^2}}\right),
\label{Gaussianapprox}
\ea
where
\ba
&m_K \equiv \frac{\left(\Omega_+ + 1/2 \right)\mathcal{H}_K}{(N\Df)\mathcal{H}_K},\\
&\sigma_K \equiv  \sqrt{ \frac{\left(\Omega_+ + 1/2 \right)^2\mathcal{H}_K}{(N\Df)^2\mathcal{H}_K} -m_K^2 }, 
\ea
are the mean and standard deviations of each saddle, with the position dependence of the higher-spin conformal blocks suppressed. 

At the level of the moment-generating function, this gives
\ba
\mathfrak{M}_K(t,z) = e^{m_K t + \frac{1}{2} \sigma^2_K t^2 + O(t^3)} \mathcal{H}_K(P,P/Q),
\ea
where the $O(t^3)$ terms correct the skew and higher moments associated with this ansatz. 

We now just need to compute the $m_K,\sigma_K$ associated with each saddle, and plug the result back into the form of the total measure
\ba
\mu(\tilde{\Delta}) = \delta(\tilde{\Delta}) + \sum^N_{K=1} \binom{N}{K}^2 \mu_K(\tilde{\Delta}).
\label{totalderivedmeasure}
\ea
We expect these Gaussian corrections for the derived measure to only be valid near the heavy limit, when the OPE distribution can be approximated as a finite sum of saddle points. Therefore, it suffices to study the moments of the terms in $\mathcal{H}_K$ which are leading in the heavy limit for Euclidean configurations, namely
\ba
\mathcal{H}_K(P,P/Q) \supset (P/Q)^K.
\ea
Note that other terms would give contributions that are exponentially suppressed and wouldn't affect any power-law correction terms computed below.

The first and second moments for these terms can be  computed exactly in 1d and are given by
\ba
\left(\frac{P}{Q}\right)^{-K} \left(\Omega_+ +\frac{1}{2} \right) \left(\frac{P}{Q}\right)^{K}=& \frac{1}{2} \left(4  \Delta _{\phi }K-1\right) (1-z)^{2  \Delta _{\phi }K} \mathcal{F}_{2\Df K}(z)+\frac{1}{2},\\
\left(\frac{P}{Q}\right)^{-K} \left(\Omega_+ +\frac{1}{2} \right)^2 \left(\frac{P}{Q}\right)^{K}=&\left(\frac{P}{Q}\right)^{-K} \left(\Omega_+ +\frac{1}{2} \right) \left(\frac{P}{Q}\right)^{K}-\frac{2  \Delta _{\phi }K \left(2  \Delta _{\phi }K+z-1\right)}{z-1},
\ea
where $\mathcal{F}_r (z) = {}_2 F_1 ( r,r,r-1/2;z)$. 

Additionally, we can use our asymptotic operators to compute the approximate result in general dimension
\ba
\left(\frac{P}{Q}\right)^{-K} \left(\tilde{\Omega}_+ +\frac{d}{2} \right) \left(\frac{P}{Q}\right)^{K}=&\,\frac{2  \Delta _{\phi } K }{\sqrt{1-z}}+ \frac{d \left(z+2 \sqrt{1-z}-2\right)}{4 \sqrt{1-z}}+ O\left(\frac{1}{\Df K }\right),\\
\left(\frac{P}{Q}\right)^{-K} \left(\tilde{\Omega}_+ +\frac{d}{2} \right)^2 \left(\frac{P}{Q}\right)^{K}=&\,\frac{4  \Delta _{\phi }^2 K^2}{1-z} -\frac{ \left(d \left(z+2 \sqrt{1-z}-2\right)+z\right) \Delta _{\phi }K}{z-1}\\&+\frac{d \left(2 z^2-d \left(z \left(z+4 \sqrt{1-z}-8\right)-8 \sqrt{1-z}+8\right)\right)}{16 (z-1)}\\&+ O\left(\frac{1}{\Df K}\right),
\ea
as $\Df K  \to \infty$.

With these moments in hand, there are a few key facts to point out. First, the moments generated by our Gaussian $\mathfrak{M}_K(t,z)$ match the leading terms we obtained from the $\delta$-distribution result. The subleading terms slightly shift and widen the leading $\delta$-distributions, and we can study the standard deviations of each saddle. At $z=1/2$ these are given by
\ba
\sigma_K = \frac{\sqrt{K}}{N\sqrt{\Delta_{\phi}}}\left(1 -\frac{d}{32 K \Df}+ O\left(\frac{1}{K^{2} \Df^{2}}\right) \right).
\ea

On the other hand, the locations of the saddles are separated by intervals of $2\sqrt{2}/N+O((N\Df)^{-1})$ in $\tilde{\Delta}$. This means that, for a fixed $N$, taking the heavy limit results in $N$ saddles that become relatively spaced apart, while fixing $\Df$ and taking the long limit $N \to \infty$ results in the saddles with $K \sim N$ overlapping and merging to form one large mass around $\tilde{\Delta} \sim \sqrt{2}$. 

Let us now consider some global properties of this collective saddle at $z = \bar{z} = 1/2$. We will first consider the limit of $\Df,N \to\infty$ with $\frac{\Df}{N^2} \gg1$. In this limit, the individual saddles given by $\mu_K(\tilde{\Delta})$ in eq.~(\ref{totalderivedmeasure}) have standard deviations of order $O\left(1/\sqrt{K\Df}\right)$ and tend to delta masses at $\tilde{\Delta} = 2\sqrt{2} \frac{K}{N}$ as $\Df \to \infty$. Since the spacing between masses is $O\left(1/N \right)$ with their individual width at most $O\left(1/\sqrt{\Df} \right)$, letting $\frac{\Df}{N^2} \gg1$ ensures that the width of each saddle is small compared to their separation. In this regime, the collective saddle is entirely characterized by the squared binomial coefficient factor $\binom{N}{K}^2$. Taking $K,N\to \infty$ with $K/N \sim 1/2$, we can approximate
\ba
\binom{N}{K}^2 = \binom{2N}{N}\left( \frac{2}{\sqrt{\pi N}} e^{-\frac{4}{N}\left(K-\frac{N}{2}\right)^2}+ O\left(\frac{1}{N^{3/2}}\right) \right).
\label{binomtogauss}
\ea
Reading off the parameters of this Gaussian, we see the distribution over $K$ has a mean of $N/2$ with standard deviation $\sqrt{N}/(2\sqrt{2})$. Since the delta masses are located at $\tilde{\Delta} =2\sqrt{2}\frac{K}{N} $, we can convert eq.~(\ref{binomtogauss}) into a distribution over $\tilde{\Delta}$ by setting $K = \frac{\tilde{\Delta}N}{2 \sqrt{2}}$. The collective saddle then reads
\ba
\mu_\mathrm{total}(\tilde{\Delta}) \sim \binom{2N}{N} \sqrt{\frac{N}{2\pi}} e^{-\frac{N}{2}(\tilde{\Delta}-\sqrt{2})^2},
\ea
which is a Gaussian centered at $\tilde{\Delta} = \sqrt{2}$ with standard deviation $\sigma_\mathrm{total} =1/ \sqrt{N}$. 

When $\Df$ is held finite with $N$ large, it still holds true that $\sigma_\mathrm{total}  = O(1/\sqrt{N})$ due to the fact that the standard deviation of an individual saddle around $K \sim N$ is $O\left(1/\sqrt{N} \right)$. Therefore, the widths of individual saddles giving dominant contributions to the total measure do not grow faster than the width controlled by the binomial factor. This demonstrates that our measure tends towards a $\delta$-distribution at $\tilde{\Delta} = \sqrt{2}$ in the limit of $N \to \infty$, giving rise to the measure which nearly saturates the lower bound of moment space, $\mu^{(-)}$. 

In general dimensions we can also study the spin distribution in more detail. First let's consider the generalized free theory moments of $J_2$ (subtracting the identity), computed as
\ba
\frac{L_{\mu/\{0\}}[J^j_2]}{L_{\mu/\{0\}}[1]} &= \left.\left( \frac{u}{v}\right)^{-\Df}\left(\Omega_-^{2} - \left(\frac{d-2}{2}\right)^2\right)^j \left(  \frac{u}{v}\right)^{\Df}\right|_{z=\bar{z}=1/2}\\
&=\left(\frac{d-1}{2}\right)_{j}(2\Df)^j+O(\Df^{j-1}).
\ea
Through this, we see that the generalized free $\left<\phi\phi\phi\phi\right>$ correlator is dominated by double-twist operators with spins distributed around $J_2 \sim (d-1) \Df$, with an average value of $\left<J_2\right> = \frac{(d-1)}{2} \Df$ after the identity operator is included.

Resumming the leading terms into a moment-generating function and taking the inverse Laplace transform gives a coarse-grained approximation of the OPE distribution of a heavy GFF over $J_2$. The result is a gamma distribution with shape parameter $k=(d-1)/2$ and scale parameter $\theta = 2\Df$:\footnote{The gamma distribution is given by $\Gamma(x; k, \theta) = \frac{x^{k-1} e^{-x/\theta}}{\theta^{k}\Gamma(k)}$ and has mean $k \theta$ and variance $k \theta^2$.}
\ba
 \Gamma(J_2; (d-1)/2, 2\Df) =  \frac{ J_2^{\frac{d-3}{2}}e^{-\frac{J_2}{2 \Delta _{\phi }}}  }{(2\Delta _{\phi })^{\frac{d-1}{2}}\Gamma \left(\frac{d-1}{2}\right)} .
\ea
We note that a similar computation applied to the leading $K=N/2$ saddle of the $\mathcal{G}_N$ correlator gives the same distribution but with $\Df \rightarrow \frac{N}{2} \Df = \frac12 \Delta_{\phi^N}$.

We can also see this structure emerge more directly by considering the generalized free theory OPE coefficients squared:\footnote{This is in the normalization of the 1st line of Table I of~\cite{Poland:2018epd}.}
\ba
\lambda_{\phi\phi [\phi\phi]_{n,\ell}}^2 &= \frac{(1+(-1)^{\ell})2^{\ell}\left(\Delta_{\phi}-\frac{d}{2} +1\right)_n^2 (\Delta_{\phi} )_{\ell+n}^2}{\ell! n!
   \left(\ell+\frac{d}{2}\right)_n (2 \Delta_{\phi}+n-d +1)_n (2 \Delta_{\phi}+2n+\ell -1)_{\ell} \left(2 \Delta
_{\phi}+n+\ell-\frac{d}{2} \right)_n}.
   \label{eq:GFFgenerald}
\ea
Given our discussion above, at the self-dual point $z=\bar{z}=1/2$ we expect that the $N=1$ GFF 4-point function $\langle\phi\phi\phi\phi\rangle$ at large $\Df$ is dominated by exchanged operators with dimension $\Delta \sim 2\sqrt{2} \Df = 2\Df + 2n + \ell$ and even spins distributed around $\ell \sim \sqrt{(d-1)\Df}$.  

We can see this structure by expanding eq.~(\ref{eq:GFFgenerald}) around the saddle at large $\Df$. It is convenient to work with the rescaled variables 
\ba
\tilde{J_2} &= J_2/\Df = \ell(\ell+d-2)/\Df, \\
\tilde{\Delta} &= \Delta/\Df = 2 + 2n/\Df + \ell/\Df.
\ea 
The expansion then gives
\ba
\lambda_{\phi\phi [\phi\phi]_{n,\ell}}^2 &G_{2\Df +2n+\ell}(1/2,1/2) \\ =&  \left(\frac{8 \tilde{J_2}^{\frac12}}{\Df^{\frac32}}\right) \left(\frac{\tilde{J_2}^{\frac{d-3}{2}} e^{-\tilde{J_2}/2}}{ 2^{\frac{d-1}{2}} \Gamma\left(\frac{d-1}{2}\right)} \right)  \left(\frac{e^{-\frac{\Df}{2} (\tilde{\Delta}-2\sqrt{2})^2}}{\sqrt{2\pi/\Df}} \right) \\
&\times\left[1+ O\left((\tilde{\De}-2\sqrt{2})^3 \Df\right)
+O\left((\tilde{\De}-2\sqrt{2})\tilde{J}_2 \right)
+O\left((\tilde{\De}-2\sqrt{2}) \right) \right].
\label{eq:GFFdistributions}
\ea
Up to an overall prefactor, the leading term is a normalized Gaussian distribution in $\tilde{\Delta}=\Delta/\Df$ centered at $2\sqrt{2}$ with standard deviation $\Df^{-\frac12}$ times a normalized gamma distribution in $\tilde{J}_2 = J_2/\Df \simeq \ell^2/\Df$ with shape parameter $k=(d-1)/2$ and scale parameter $\theta = 2$. The prefactor $8\tilde{J_2}^{\frac12}\Df^{-\frac32}$ exactly compensates for the change in measure from summing over even integer spaced $\Delta$ and $\ell$ to integrating over $\tilde{\Delta}$ and $\tilde{J_2}$. Thus, the double-twist operators give a total contribution of $1$ to the correlator which balances against the identity operator contribution which is also $1$. One can also easily see the emergence of the anticipated $\delta(\tilde{\Delta} - 2\sqrt{2})$ distribution by taking the $\Df \to \infty$ limit of the Gaussian.

The corrections to (\ref{eq:GFFdistributions}) start giving an $O(1)$ modification in the coefficient of the exponentials when $(\tilde{\Delta}-2\sqrt{2}) \gtrsim \Df^{-1/3}$ or when $(\tilde{\Delta} - 2\sqrt{2})\tilde{J_2} \gtrsim 1$. However, in these regimes the contributions to the correlator are exponentially suppressed at large $\Df$, so the leading term in (\ref{eq:GFFdistributions}) remains a very good approximation. Note that we are organizing the corrections assuming that $(\tilde{\Delta}-2\sqrt{2}) = O(\Df^{-1/2})$, i.e. that it is within the non-suppressed region of the Gaussian distribution.

Let us now connect back to the general covariance bounds we derived earlier in eq.~(\ref{covbound}). We can easily compute the covariance between $\Delta$ and $J_2$ in the $\mathcal{G}_N$ correlators by applying the appropriate shifted $\Omega$-operators. At leading order in large $N\Df$, we find 
\ba
\mathrm{Cov}(\tilde{\Delta},\tilde{J_2})[ \mathcal{G}_N(z = \bar{z}=1/2)] = \frac{d-1}{(2N-1)\sqrt{2}}\left(1 + O\left(\frac{1}{N\Df}\right)\right).
\ea
We notice at $N=1$, the asymptotic upper bound given by eq.~(\ref{covbound}) is saturated, while for $N \to \infty$, the lower bound is saturated. Therefore, not only does $\mathcal{G}_1$ saturate the upper bound of scaling moment space as $\Df \to \infty$, but it also maximizes the positive correlation between angular momentum and scaling dimension in the OPE distribution. On the other hand, the $\mathcal{G}_N$ correlator has an OPE distribution such that $\Delta$ and $J_2$ are minimally correlated in the long limit $N \to \infty$. 

The extreme values of the covariance bound can also be understood as arising from the extremal distributions $\mu^{\pm}(\tilde{\Delta}, \tilde{J_2})$ in both scaling dimension and spin which account for the gamma distribution in $\tilde{J_2}$. Concretely, the asymptotic measures discussed in section~\ref{sec:saddles} can be generalized to
\ba
\mu^{(+)}(\tilde{\Delta},\tilde{J_2})= \frac{1}{2}\left( \delta(\tilde{\Delta})\delta(\tilde{J_2}) + \delta(\tilde{\Delta}-2\sqrt{2}) \Gamma\left(\tilde{J_2}; (d-1)/2, 2\right) \right)
\ea
and
\ba
\mu^{(-)}(\tilde{\Delta},\tilde{J_2}) = \delta(\tilde{\Delta}-\sqrt{2}) \Gamma\left(\tilde{J_2}; (d-1)/2, 1\right).
\ea
These approximate the large-$\Delta_{\phi}$ asymptotics of the $\mathcal{G}_1$ correlator and the large-$N$ asymptotics of the $\mathcal{G}_N$ correlator, respectively, up to their overall normalization. 

They correspond to the 2-variable moment generating functions
\ba
M^{(+)}_{\tilde{\Delta},\tilde{J_2}}(t,s) = \frac{1}{2}\left(1+ \frac{e^{2\sqrt{2} t}}{(1-2 s)^{\frac{d-1}{2}}}\right)
\ea
and
\ba
M^{(-)}_{\tilde{\Delta},\tilde{J_2}}(t,s) = \frac{e^{\sqrt{2} t}}{(1-s)^{\frac{d-1}{2}}}.
\ea
We have verified that these reproduce the leading asymptotic behavior of the mixed moments of the GFF correlators computed using the $\tilde{\Omega}_{\pm}$ operators.

\subsection{Weight-interpolating functions}
\label{sec:WIFs}

In section \ref{sec:gffsaddles}, we used Gaussians to model some of the properties of general saddles based on their mean, variance, and normalization. This choice is made for two main reasons: 
\begin{itemize}
    \item A Gaussian is the maximum entropy distribution for a fixed mean and variance, making it the ``simplest" distribution that matches those low-lying moments.
    \item Up to a determined factor, our derived Gaussian measures converge uniformly to the exact weights of local operators in the spectrum of each GFF saddle in the heavy limit.
\end{itemize}
Placing the second bullet on a more rigorous footing will be the focus of this section, and we begin with a definition.

Consider the weighted OPE distribution over scaling dimension at $z = \bar{z} = z^\star$,
\ba
\mu^\star(\Delta, J_2)  = \delta(\Delta)\delta(J_2) + \sum_{\Delta'>0,\ell'} \delta(\Delta-\Delta')\delta(J_2-J_2')a_{\Delta',\ell'} G_{\Delta',\ell'}(z^\star).
\ea
A weight-interpolating function (WIF) satisfies
\ba
I(\Delta,J_2; z^\star) = a_{\Delta,\ell}G_{\Delta,\ell}(z^\star)
\ea
for all $\Delta,J_2 \equiv\ell(\ell+d-2)$ in the discrete support of $\mu^\star(\Delta,J_2)$. Such a function is not unique, and one can be directly constructed from the measure as
\ba
I(\Delta,J_2; z^\star) = \int_{J_2-\epsilon}^{J_2 + \epsilon}\int_{\Delta - \epsilon}^{\Delta+\epsilon} d\Delta' dJ_2' \;\mu^\star(\Delta',J_2')
\ea
for all $2\epsilon$ less than the difference in scaling dimension and total angular momentum between any two operators in the OPE. The resulting WIF is not generically smooth or continuous. If the OPE spectrum is uniformly spaced in $k$, then it is possible to construct a (piecewise) linear WIF over scaling dimension by smearing the OPE measure over an appropriate kernel. 

Let us explicitly construct this linear interpolation function for the weights of an equally spaced discrete `target' distribution $\rho(x) = \sum_n a_n \delta(x-\kappa n)$ where $\kappa$ is the spacing between each $\delta$-distribution and $a_n$ are some positive weights. $\rho(x)$ need not be normalized. A linear interpolating function for this distribution should satisfy $I( \kappa n) = a_n$ for all $n$ and $I(\kappa n p + \kappa (n+1) (1-p) )=  a_n p + a_{n+1} (1-p)$ for $p \in (0,1)$. 

Such a function can be obtained by convolution with a triangle function
\ba
I(x) =\Lambda\ast\rho =\int_\mathbb{R} \Lambda\left(\frac{x-t}{\kappa}\right)\rho(t) dt,
\ea
where
\begin{align}
\Lambda(x) \ 
   &= \begin{cases}
      1 - |x|, & |x| < 1; \\
      0        & \text{otherwise}. \\
      \end{cases}
\end{align}
If we were to compute the moments of $I(x)$, we would see:
\ba
\int_\mathbb{R} x^j I(x) dx = \kappa \int_\mathbb{R}\left( t^j +O(t^{j-2})  \right)\rho(t) dt ,
\ea
or that the moments of the linear interpolation function are approximately those of the target distribution multiplied by the spacing between $\delta$-distributions, up to a correction by a sub-subleading moment. In the context of our problem, where moments are organized in an expansion around $\Df \to \infty$, this implies $\nu^I_k - \kappa \nu_k  = O(\Df^{k-2})$, where $\{\nu^I_k\}_{k \geq0}$ is the moment sequence of an approximately linear WIF. 

This property is also satisfied by the derived Gaussian measure we used to study the OPE distribution in GFF correlators when operators in the spectrum are separated by $\kappa = 2$. Thus, the total derived measure
\ba
 \mu^\star_{(G)}(\Delta) = \frac{2}{N\Df} \sum_{K=1}^N \binom{N}{K}^2 \mu^\star_K(\Delta/(N\Df)) 
\ea
satisfies the properties of a WIF in the heavy limit, where $\mu^\star_K(\tilde{\Delta})$ is the Gaussian measure in eq.~(\ref{Gaussianapprox}) evaluated at $z= \bar{z} = z^\star$. In addition to checking this agreement graphically in a number of examples, we also prove uniform convergence for the simple case of $N=1$ and $z^\star = 1/2$ in $d=1$, or when the target weights are known as a simple analytic function for $\Delta > 0$, the derived measure is a single Gaussian, and the normalization rapidly approaches $1$ in the heavy limit. To condense notation, we will adopt the convention of $I(\Delta) \equiv I(\Delta;1/2)$.

We say $I(\Delta)$ uniformly converges to the exact weights $a^{GFF}_n[\Df] G_{2\Df + 2n}(1/2)$ if for every $\epsilon > 0$ there exists a $\Df'$ such that for all $\Df\geq\Df'$ and $n\in\mathbb{N}$ 
\ba
|a^{GFF}_{n}[\Df] G_{2\Df+2n}(1/2) - I(2\Df+2n)| <\epsilon.
\ea
To derive asymptotics in the heavy limit, let us re-parametrize by setting $n =  (\sqrt{2}-1)\Df + \frac{\delta}{2}\sqrt{\Df}$ where $\delta$ parametrizes the number of standard deviations (of order $\sqrt{\Df}$) one is from the mean of the leading saddle. Plugging this in and expanding around $\Df \to \infty$ gives
\ba
a^{GFF}_{n}[\Df] G_{2\Df+2n}(1/2) = e^{-\frac{\delta^2}{2}}\left( \sqrt{\frac{2}{\pi \Df}} -\frac{\delta  \left(4 \delta ^2+6 \sqrt{2}-21\right)}{12 \sqrt{\pi } \Df} + O\left(\frac{1}{\Df^{3/2}} \right) \right)
\ea
and
\ba
I(2\Df+2n) = e^{-\frac{\delta^2}{2}}\left( \sqrt{\frac{2}{\pi \Df}} +\frac{2 \sqrt{2}-3}{4 \sqrt{\pi }}\frac{ \delta }{\Delta _{\phi }} + O\left(\frac{1}{\Df^{3/2}} \right) \right).
\ea
Subtracting these results and bounding the difference gives
\ba
|a^{GFF}_{n}[\Df] G_{2\Df+2n}(1/2) - I(2\Df+2n)| \leq \frac{|\Upsilon(\delta) | }{\Df},
\ea
 where 
\ba
\Upsilon(\delta) = -\frac{e^{-\frac{\delta ^2}{2}} \delta  \left(2 \delta ^2+6 \sqrt{2}-15\right)}{6 \sqrt{\pi }} \approx e^{-\frac{\delta ^2}{2}} \left(0.612589 \delta -0.188063 \delta ^3\right)
\ea
takes its maximal absolute value at $\delta \approx \pm 0.756996$ so that
\ba
|\Upsilon(\delta)| \leq 0.286944.
\ea
Since $|\Upsilon(\delta)|$ is bounded by a constant for all $\delta \equiv \frac{2 n}{\sqrt{\Df }}+2\left(1- \sqrt{2}\right) \sqrt{\Df }$, and therefore all $n$, $I(\Delta)$ converges uniformly to the exact weights in the heavy limit with errors of order $\Df^{-1}$. 

\begin{figure}[tbp]
\centering
\includegraphics[width=\textwidth]{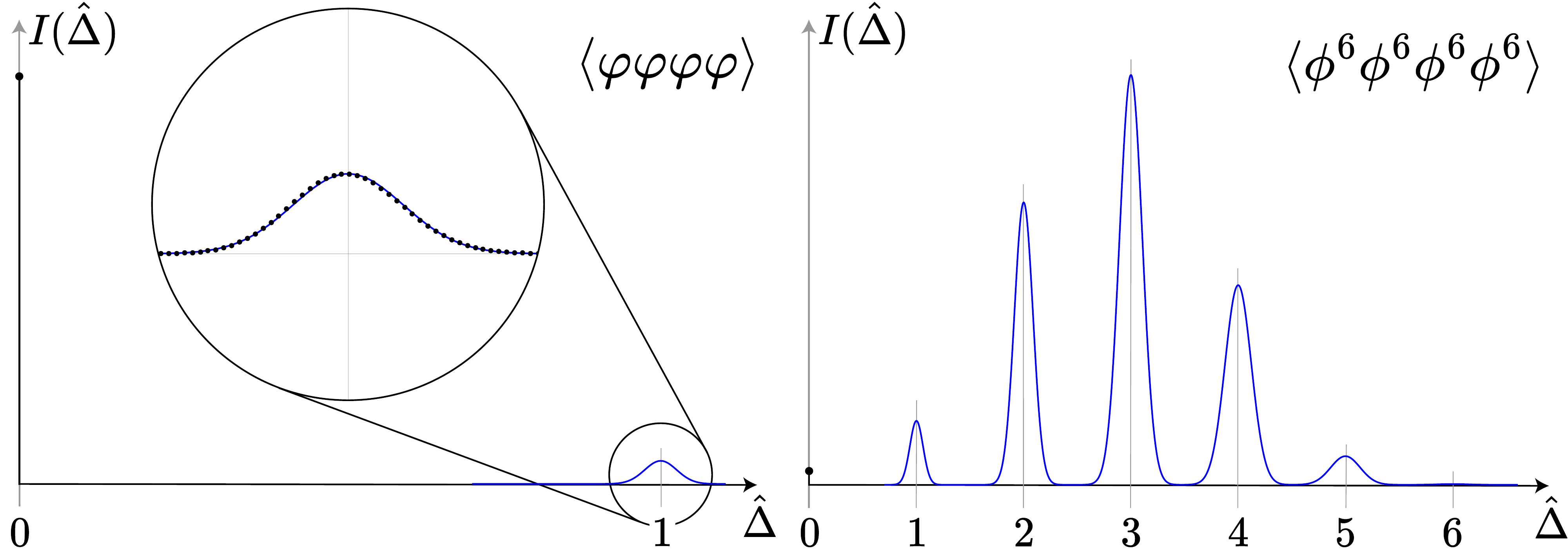}
\caption{Weight-interpolating functions at $z=1/2$ for correlators in generalized free field theories with $\Delta_\varphi = \Delta_{\phi^6} = 200$. The $x$-axis is labeled by a rescaled scaling dimension $\hat{\Delta} = \Delta/\left(2\sqrt{2}\Delta_{\phi/\varphi}\right)$, so that saddles are located at $\hat{\Delta} = K$. The black dots mark the exact weights of operators in the OPE distribution, including the $s$-channel identity at $\De=0$. The saddle of largest scaling dimension is associated with the t-channel identity.}
\label{freewif}
\end{figure}

To give some examples, in the LHS of fig.~\ref{freewif} we plot the exact weights of double-twist operators in the $\mathcal{G}_1(1/2)$ OPE for a large external scaling dimension $\Delta_{\varphi} = 200$ against the Gaussian WIF we computed from its moments. In the RHS, we plot the WIF for the $\mathcal{G}_6(1/2)$ OPE expressed as a sum over Gaussians, tuning the scaling dimension of a single field $\phi$ such that $\Delta_{\varphi} = \Delta_{\phi^6} = 200$.

\subsection{Bulk contact interactions}
Given a correlator, one would like to extract the underlying CFT data, which enables the calculation of critical exponents and provides holographic insights into bulk physics. Light correlators tend to be governed by a small number of light states, whereas heavy correlators are dominated by numerous heavy states. This complexity of the high-dimension spectrum poses a challenge, especially in heavy perturbative correlators, where unmixing the CFT data remains difficult even when using the Lorentzian inversion formula. These challenges become more pronounced in holographic theories with non-renormalizable bulk interactions, where heavy states are highly sensitive to the UV behavior. 

The approach we offer to gain insights into this challenging physics is to treat the CFT data as a coarse-grained distribution over scaling dimensions and spin, and examine how interactions affect the descriptive statistics of this smooth distribution rather than focusing on a few discrete data points. By utilizing the basis of HS conformal blocks and their associated OPE distributions, we can gain new perspectives into the physics of heavy CFT correlators, dual to the bulk physics of heavy states, advancing our understanding of quantum many-body physics in gravitational theories.

The coarse-grained OPE distributions we compute offer not only qualitative insights but can also be useful quantitatively. In section \ref{sec:WIFs}, we showed that rescaled Gaussian measures, or WIFs, converge uniformly to the exact weights of the GFF spectrum in the heavy limit. In the interacting case, Gaussian WIFs computed from the perturbed data remain highly accurate approximations of the exact weights in the heavy limit, even at finite coupling. We dub this phenomenon ``Gaussianization" and verify that saddles perturbed by bulk contact diagrams with an arbitrary number of derivatives Gaussianize as $\Df \to \infty$.

Let us consider perturbing the AdS bulk action~(\ref{bulkGFFaction}) by a contact interaction containing $2L$ derivatives:
\ba
S^{\text{int}} = g_L \int_{\text{AdS}}  (\phi \partial^L \phi)^2.
\label{cinteraction}
\ea
This interaction has been extensively studied at tree level in AdS${}_2$~\cite{Bianchi2021-li,Knop2022-qh}, and the anomalous dimensions of double-twist families with scaling dimension $\Delta_n = 2 \Df + 2n + g_L \gamma_{L,n}^{(1)}$ have been computed in closed form to be
\ba
\gamma_{L,n}^{(1)}&= -\frac{2^{-2 \Delta _{\phi }-1} \Gamma \left(\frac{L}{2}+\Delta _{\phi }\right) \Gamma \left(L+2 \Delta _{\phi }-1\right) \Gamma \left(\frac{3 L}{2}+2 \Delta _{\phi }-\frac{1}{2}\right)}{\sqrt{\pi } \Gamma \left(L+\Delta _{\phi }-\frac{1}{2}\right)} \\
&\times\frac{\Gamma \left(n+\frac{1}{2}\right) \Gamma \left(n+\Delta _{\phi }\right) \Gamma \left(-\frac{L}{2}+n+\Delta _{\phi }\right) \Gamma \left(\frac{L}{2}+n+2 \Delta _{\phi }-\frac{1}{2}\right)}{\Gamma \left(-\frac{L}{2}+n+1\right) \Gamma \left(n+\Delta _{\phi }+\frac{1}{2}\right) \Gamma \left(n+2 \Delta _{\phi }\right) \Gamma \left(\frac{L}{2}+n+\Delta _{\phi }+\frac{1}{2}\right)}\\
&\times \, _4\tilde{F}_3\left(-L,-n,2 \Delta _{\phi }+L-1,2 \Delta _{\phi }+n-\frac{1}{2};\Delta _{\phi },\Delta _{\phi }-\frac{L}{2},2 \Delta _{\phi }+\frac{L}{2}-\frac{1}{2};1\right).
\ea
This formula admits a simple asymptotic form as $n,\Df \to \infty$ with $n/\Df$  held constant, giving
\ba
\gamma_{L,n}^{(1)}  \sim -\frac{1}{\pi}2^{2 L-3} n^{\frac{L-1}{2}} \left(\Delta _{\phi }+n\right){}^{L-1} \left(2 \Delta _{\phi }+n\right){}^{\frac{L-1}{2}}.
\ea

This asymptotic behavior will be sufficient for our analysis of the OPE measure at $z = 1/2$, since the OPE is dominated by double-twist operators with $n \sim (\sqrt{2}-1)\Df$ as $\Df \to \infty$. Note that for $L=0$, this interaction is a relevant operator in an AdS${}_2$ bulk, so anomalous dimensions vanish as $n \to \infty$~\cite{Fitzpatrick2010-wk}. This means that heavy saddles are robust to perturbations by this operator and remain well-approximated by the free theory result. On the other hand, the operator for $L>1$ is irrelevant, so anomalous dimensions grow with $n$. This means there are non-trivial deformations on heavy saddles which we can measure by studying how moments are shifted in the presence of the interaction. The marginal case of $L=1$ results in $\gamma_{n,1}$ being a constant, and saddles are merely shifted by an amount proportional to $g_1$. In all cases, for  $n = (\sqrt{2} -1)\Df$ the anomalous dimensions go as $\gamma_{L,n} \sim -\frac{1}{\pi} 2^{\frac{5 L}{2}-\frac{7}{2}} \Delta _{\phi }^{2 L-2}$, therefore we can take $g_L = g/\Df^{2L-2}$ to cancel out the large $\Df$ dependence so that $g_L \gamma_{L,n} = O(g)$ for all $\Df$.

Neglecting the $O(g^2)$ corrections, we can compute the moments explicitly as
\ba
\nu_k = \sum_n (a^{(0)}_{n} + g_L a^{(1)}_{L,n}) (2\Df+ 2n + g_L \gamma^{(1)}_{L,n})^k G_{2\Df+ 2n + g_L \gamma^{(1)}_{L,n} }(1/2),
\label{momentsum}
\ea
 where the anomalous OPE coefficients of the double twist operators are computed using the derivative rule of $a_{L,n}^{(1)} = \frac{1}{2}\partial_n\left(a_{L,n}^{(0)} \gamma_{L,n}^{(1)}\right)$ from \cite{Heemskerk2009-mb}. This procedure may be thought of as a resummation of tree-level data into the moment variables, with deviations from the true all-loop order moments arising at order $g^2$. In practice, eq.~(\ref{momentsum}) is computed by summing over operator contributions in a large window around the saddle at $n \sim (\sqrt{2} - 1)\Df$. We then plug these moments into the Gaussian ansatz in eq.~(\ref{Gaussianapprox}) evaluated at $z= 1/2$, and multiply by the appropriate $\kappa$ factor to produce our desired perturbed WIF. 

\begin{figure}[tbp]
\centering
\includegraphics[width=\textwidth]{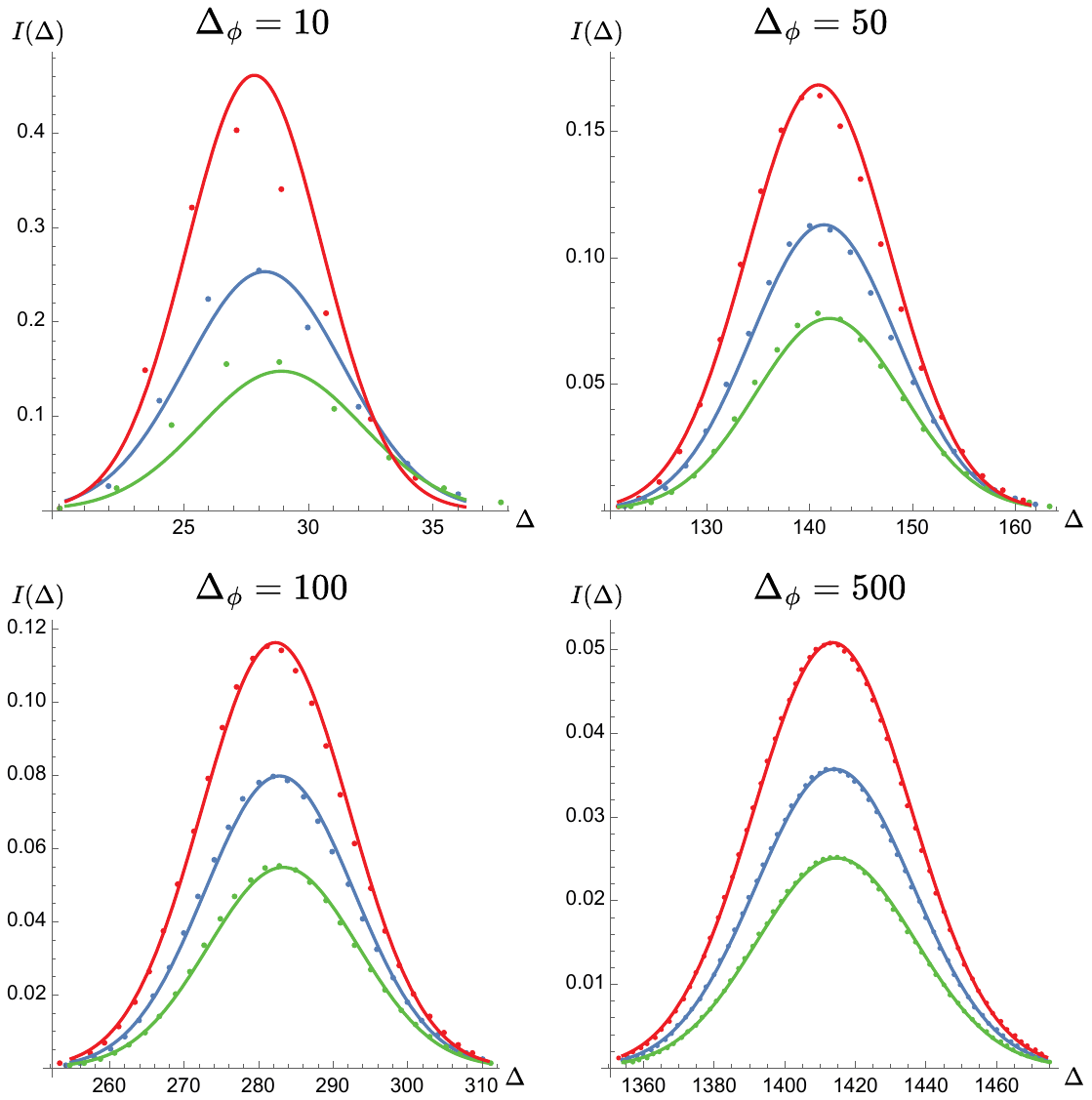}
\caption{Gaussian weight-interpolating functions (WIFs) at $z= 1/2$, plotted against the exact weights of operators in $\mathcal{G}_1$ coupled to the interaction (\ref{cinteraction}) with $L=2$ at different values of $\Df$ and $g$. Blue: $g=0$, Green: $g = -1$, Red: $g=1$.}
\label{perturbedWIF}
\end{figure}
 
 We can approximate $\kappa$ by considering the spacing between operators around the saddle point, i.e.~when $n \sim \eta \Df$ for $\eta = O(1)$. Let $\kappa_n = \Delta_{n+1} - \Delta_{n}$. Taking $n = \eta \Df$ and $\Df \to \infty$, we find
\ba
\kappa_{\eta \Df} &= 2 + g_L \left( \gamma_{L,\eta \Df+1} -  \gamma_{L,\eta \Df} \right)\\
&= 2 + g \frac{(2 \eta  (\eta +2)+1) 2^{2 L-3} (L-1) (\eta +1)^{L-2} (\eta  (\eta +2))^{\frac{L-3}{2}}}{\pi  \Delta _{\phi }}.
\ea
The numerator of the anomalous term is $O(1)$ for operator spacings around the saddle, so $\kappa = 2 + O(1/\Df)$. Neglecting this error term, we can simply set $\kappa = 2$ to obtain our perturbed WIFs in the heavy limit.

We find that the perturbed WIFs computed using the Gaussian approximation do an excellent job of capturing the shape of the spectrum at all perturbative values of $g$, and even at $O(1)$ values. To illustrate this, in fig.~\ref{perturbedWIF} we plot these perturbed WIFs against the spectrum of $\mathcal{G}_1$ with an $L=2$ contact interaction for different values of the coupling and external scaling dimension. We present a sequence of plots with $\Df = 10,50,100,500$ to show how both free and interacting spectra tend towards Gaussian WIFs in the heavy limit. The required moments were directly computed by summing over a window of 120 operators around $n = (\sqrt{2} -1)\Df$, capturing the contributions of operators within $\gtrsim 5$ standard deviations from the mean. 

 \begin{figure}[tbp]
\centering
\includegraphics[width=\textwidth]{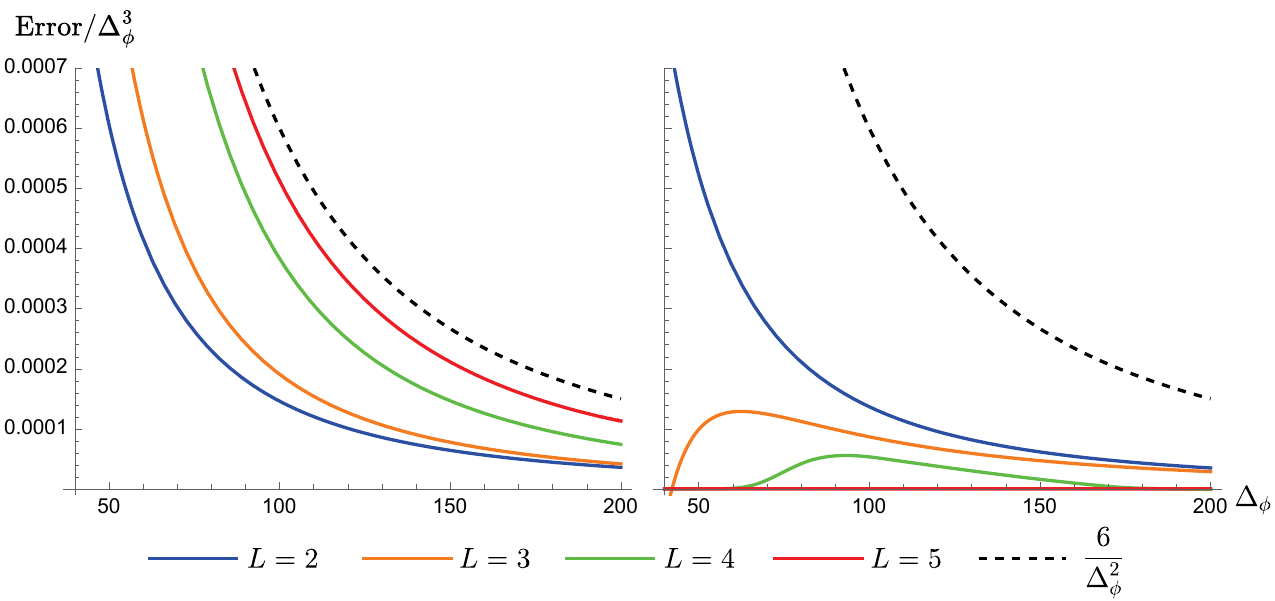}
\caption{The regulated Error $=\frac{\nu_k}{\nu_0} -\left.\partial^k_t e^{\mu t + \frac{\sigma^2}{2} t ^ 2 }\right|_{t = 0}$ away from Gaussian moments for $k=3$, plotted as a function of $\Df$. The left plot is computed with $g=-1$, while the right plot is computed with $g= 1$. The dashed line is a reference bound showing that saddles in either interaction tend to Gaussianize with errors of order $\Df^{k-2}$.}
\label{gaussianization}
\end{figure}

 One way to quantitatively test the ``Gaussianity" of the perturbed OPE distribution is by comparing its exact higher moments to those predicted by the Gaussian ansatz. Namely, we can ask whether
 \ba
 \frac{\nu_k}{\nu_0} -\left.\partial^k_t e^{\mu t + \frac{\sigma^2}{2} t ^ 2 }\right|_{t = 0} = O(\Df^{k-2})
 \ea
 as $\Df \to \infty$ for all $k>2$. If this condition is satisfied, then we can reconstruct the WIF for a given perturbed OPE density from the first two exact moments as a Gaussian in the heavy limit, with corrections arising at sub-subleading order in $\Df$. 
 
 In fig.~\ref{gaussianization}, we check these error terms for the $\nu_3/\nu_0$ moment of a $\mathcal{G}_1$ correlator perturbed by contact interactions with $2L$ derivatives, and find that this condition is indeed satisfied for all $L$ we were able to feasibly check. Perhaps more interestingly, the rate at which the OPE distribution associated with a different interaction Gaussianizes is dependent on the sign of the coupling. We find that higher-derivative interactions Gaussianize slower with a negative coupling, while lower-derivative interactions Gaussianize slower with a positive coupling. A more in-depth analytical investigation of this saddle Gaussianization would be extremely interesting, as would tests of Gaussianization at higher order in the coupling.

\section{Discussion}
\label{sec:discussion}

In this paper, we have proposed the use of classical moments in $\Delta$ and $J_2 \equiv \ell(\ell+d-2)$ as a useful way of repackaging CFT data, focusing on applications for ``heavy" correlators of identical scalar operators with $\Df \gg \frac{d-2}{2}$. This analysis is dependent on the unitary OPE being encoded by a positive definite measure over scaling dimension and total angular momentum, with the full correlator viewed as a moment-generating function for the OPE distribution. 

The latter construction relies on the existence of operators which extract the necessary powers of $\Delta$ and $J_2$ from the conformal block, which we construct exactly with Riemann-Liouville-type fractional derivative operators in $d = 1, 2,$ and $4$, and construct asymptotically with integer-derivative operators for conformal blocks of large scaling dimension in general dimension. The exact operators $\Omega_\pm$ make use of the transformation introduced by~\cite{song2023compactform3dconformal}, which we dress with an additional factor so that it acts naturally on the prefactor of the 4d conformal block. These operators allow us to easily generate moments using the action
\ba
\left(\Omega_+ +\frac{d}{2} \right)^m \left( \Omega_-^2 - \left(\frac{d-2}{2}\right) \right)^n G_{\Delta,\ell} =\Delta^m J_2^n G_{\Delta,\ell}.
\ea
It would be interesting in future work to construct exact $\Omega_{\pm}$ operators in general dimensions, as well as their generalizations to mixed correlators and higher-point functions. Such operators would give us even more powerful tools for studying the statistics of CFT data.

Applying powers of $\left(\Omega_+ +\frac{d}{2} \right)$ and $ \left( \Omega_-^2 - \left(\frac{d-2}{2}\right)\right)$ to the correlator produces a kinematic-dependent double moment sequence given by
\ba
\left(\Omega_+ +\frac{d}{2} \right)^m \left( \Omega_-^2 - \left(\frac{d-2}{2}\right)\right)^n \mathcal{G}(z,\bar{z}) &= \sum_{\Delta,\ell} a_{\Delta,\ell}  \Delta^m J_2^n G_{\Delta,\ell}(z,\bar{z})\\
&= \nu_{m,n}(z,\bar{z}).
\ea
When evaluated at $z = \bar{z} = 1/2$, we prove that the moment sequence $\nu_{m,n} \equiv \nu_{m,n}(1/2,1/2) $ for $m,n \in \mathbb{N}$ is a determinant solution to the Stieltjes double moment problem, and therefore uniquely determines the underlying OPE measure defined as 
\ba
\mu(\Delta,J_2) = \delta(\Delta)\delta(J_2) + \sum_{\Delta'>0,\ell'} a_{\Delta',\ell'} G_{\Delta',\ell'}(1/2,1/2)\delta(\Delta - \Delta') \delta(J_2 - J_2').
\ea

We can use crossing symmetry to constrain moments by Taylor expanding the crossing equation around the diagonal self-dual point $z = \bar{z} = 1/2$ and imposing that the coefficients vanish at each order. In the limit of large $\Df$, we may use the asymptotic conformal block to derive these constraints, and we obtain polynomial relations between moments at each finite derivative order. Combining these constraints with a lower bound arising from the Hankel matrix positivity of the moment sequence, we find that the leading constraint at large $\Df$ organizes into a simple constraint on crossing-symmetric OPE distributions in the heavy limit, posed as the vanishing of odd central moments:
\ba
\int_0^\infty d\Delta \;d J_2\;\mu(\Delta,J_2)\left(  (\Delta-\sqrt{2}\Df)^{2n +1} + O(\Delta^{2n}) \right) = 0 \quad \forall n \in \mathbb{N}.
\label{reflectionsymmetry}
\ea

This relation was previously explicitly derived in appendix D of~\cite{Paulos2016-mh} to study the flat space limit of AdS, and is a restriction of an approximate ``reflection symmetry" of the OPE~\cite{Kim2015-sg} to the diagonal self-dual point of $z = \bar{z} = 1/2$. We combined the constraint of (\ref{reflectionsymmetry}) with Jensen's inequality to derive two-sided bounds on the leading large $\Df$ behavior of normalized moments in $\Delta$:
\ba
2^{n/2}\leq \frac{\nu_{n,0}}{\nu_{0,0}} \Df^{-n}+O\left(\Df^{-1} \right) \leq 2^{3n/2-1}.
\label{leadingbound}
\ea

While this is a novel result in the study of classical moment sequences of correlators, a seemingly related bound was proposed in~\cite{Sen_2019}  (see eq.~(5.5)), where geometric ``moment" methods were used to derive a window in the OPE $\sim \sqrt{2} \Df < \Delta < 2\sqrt{2} \Df$ guaranteed to include at least one primary operator. It is debatable as to which one of these bounds is ``stronger." On the one hand, the authors of~\cite{Sen_2019} derived a rigorous statement about the presence of operator(s) in this window, but it does not give information about where operators may be clustered in this window or which operators are contributing most to the OPE. While our bound may not constrain the locations of operators in an exact sense, it does make a strong statement that operators should be dominantly distributed in the OPE around $\sim\sqrt{2} \Df$ with a maximum variance of $\sim 2\Df^2$ , demonstrating how the collective behavior of operator contributions is constrained by the bootstrap. In addition, the two extremal solutions saturating our bounds contain non-identity saddles at $\sqrt{2} \Df$ and $2\sqrt{2} \Df$, respectively. In this sense, we view these bounds as complimentary -- one proving the existence of individual operators in this window, and the other proving that operators must collectively cluster in this window and dominate the OPE. 

We also note that the methods used in~\cite{Sen_2019,Huang2019TheGO,Arkani_Hamed_2019} are qualitatively similar to ours. Namely, they introduce the sequence of moments given by the Taylor coefficients of the correlator around the diagonal self-dual point. This choice certainly has its benefits, in that crossing can be understood as restricting truncated moment sequences to a hyperplane in the projective moment space. Additionally, one does not require the kind of fractional derivative operators we used to obtain the moments of a correlator. What this method may lack, however, is a more direct interpretation of each of the moments in terms of CFT data. This makes it difficult to go from the simple (and exact) constraints from crossing to compelling statements about OPE data which extend those produced by the numerical bootstrap. 

Our method has countering strengths. While it is difficult to analytically derive exact constraints on classical moments from crossing, the relations we are able to derive can be directly interpreted as bounds on descriptive statistics of CFT data and give insights into the global structure of the OPE. It would be interesting to further unify our results with those presented in~\cite{Sen_2019,Huang2019TheGO,Arkani_Hamed_2019} by constructing an explicit mapping between the classical and geometric moment basis along with their relations from crossing. In upcoming work, we plan to extend our analytic study to use semidefinite optimization methods to exactly constrain classical moments, augmenting bounds produced by the standard numerical bootstrap by giving new quantitative insights into the contributions of high-dimension CFT operators. 

In addition to deriving a constraint equation for subleading terms of moments in the heavy limit (restricting to diagonal kinematics), we computed a relation between moments in the spin Casimir and scaling dimension and combined them with Hankel matrix positivity to obtain a two-sided bound on the leading term in the covariance
\ba
0 \leq \frac{\text{Cov}(\Delta,J_2)}{\Df^2} + O\left(\Df^{-1}\right) \leq \frac{d-1}{\sqrt{2}}.
\label{genunitarity}
\ea
This is an intriguing result that can be thought of as an ``averaged" unitarity bound on the behavior of heavy spinning operators in scalar correlators. The standard unitarity bound for spinning operators states that $\Delta \geq \ell+ d-2$. This naturally suggests that in a unitary OPE we should expect heavy operators to be correlated with operators with higher spin. The lower bound in eq.~(\ref{genunitarity}) confirms this fact, and the upper bound additionally states that there is a universal bound on the rate at which average scaling dimension grows with average spin. In future work, we plan to probe this bound in kinematic configurations away from the diagonal self-dual point, focusing on Lorentzian configurations where the OPE may become dominated by larger spin contributions. Such bounds may be useful in understanding the distribution of operators over spin along a given Regge trajectory.

After constraining the allowed moment space for unitary and crossing-symmetric correlators of identical scalars, we wanted to understand where interesting solutions to crossing lie in this moment space, and how one can reconstruct the OPE distribution of a correlator given its low-lying moments.  We first computed the ``extremal" measures which have moment sequences that saturate the upper and lower bounds of eq.~(\ref{leadingbound}), and found they are given by saddle point solutions with equally weighted $\delta$-distributions at $0$ and $2\sqrt{2}\Df$ for the maximal case, and a single $\delta$-distribution at $\sqrt{2}\Df$ for the minimal case. These asymptotic solutions to crossing can be obtained by taking different limits of the $\mathcal{G}_N = \langle \phi^N \phi^N \phi^N \phi^N\rangle$ correlator in a GFF. Namely, the maximal solution is obtained by taking $\Df \to \infty$ with $N=1$, and the minimal solution is obtained by taking $N \to \infty$. In the latter case, the operator families coalesce into a single collective saddle peaked around $\sqrt{2}\Delta_{\phi^N}$. On the other hand, for $\Df \to \infty$ and finite $N$, we find that the OPE distribution over scaling dimension for this correlator is approximated by $N$ non-identity saddle points distributed symmetrically around $\sqrt{2}\Delta_{\phi^N}$ at the locations $2\sqrt{2}K\Df$ for $K \in [1,N]$. Each one of these saddles is associated with a ``higher-spin" (HS) conformal block, first introduced in~\cite{Alday2016-kz}, which repackages operator families of fixed length $2K$, and are holographically dual to multi-parton states in the AdS bulk. 

The approximate locations of saddles associated with the HS block decomposition were identified by studying moments of the correlator at leading order in the $\Df \to \infty$ limit. As one might expect, subleading terms correct these locations and give saddle points a finite width. By truncating computed moments at subleading order in large $\Df$, the derived measures associated with each saddle become Gaussian distributions with widths of order $\sqrt{\Df}$. A similar effect within the Mellin space decomposition of a tree-level Witten diagram was observed by the authors of~\cite{Paulos2016-mh} in a study of the flat space limit of AdS. When the flat space limit of $R \to \infty$ was taken, the sum over power laws with Gaussian weights sharpened into a $\delta$-distribution identified with a single massive pole in the flat space S-matrix. The ``Gaussianization" we observe for each saddle in the OPE is a larger collective effect, where a sum over conformal blocks with Gaussian weights appear to coalesce into $\delta$-distributions in the $\Df \to \infty$ limit. 

Up to a factor determined by the operator spacing of the spectrum, these Gaussian approximations for the OPE measure tend to interpolate the exact weights of operators in a given saddle, converging uniformly with errors of order $\frac{1}{\Df}$ as $\Df \to \infty$. We verified this analytically in the free theory, and graphically for theories with an irrelevant bulk contact interaction (focusing on the t-channel identity saddle in the $\mathcal{G}_1$ correlator). In both cases, the weight-interpolating functions (WIFs) we derive provide quantitative predictions for OPE coefficients as a function of scaling dimension in a given saddle, using only the first three moments (including the normalization) for each saddle. This illustrates a somewhat uncanny ability of the low-lying moment variables to capture the CFT data of high-dimension operator contributions to the OPE. The caveat of this technique is that the actual locations of individual operators are lost, and the same WIF applies to multiple unique spectra. That said, in correlators whose high-dimension spectrum becomes nearly dense, these WIFs seem to be the best way of predicting this non-universal data as a ``coarse grained" description of a large number of operators. It would be useful to study the effects of higher moments such as the skew and kurtosis on these WIFs, and see how precisely one can interpolate the exact weights of the conformal block decomposition. It might also be interesting to formulate a general numerical bootstrap program for heavy correlators by studying crossing constraints on an ansatz written as a sum of (nearly) Gaussian WIFs.

\acknowledgments

We thank Li-Yuan Chiang, Murat Kolo\u{g}lu, Petr Kravchuk, Tony Liu, Dalimil Maz\'a\v{c}, Alessio Miscioscia, Matthew Mitchell, Ian Moult, Sridip Pal, Riccardo Rattazzi, Marten Reehorst, Petar Tadi\'c, Yuan Xin, and Xiang Zhao for discussions. The authors were supported by Simons Foundation grant 488651 (Simons Collaboration on the Nonperturbative Bootstrap) and DOE grant DE-SC0017660.

\newpage
\begin{appendix}

\section{Mathematical results for OPE moments}
\label{opemomentsproofs}
Let
\ba
\mu(\Delta,J_2) = \delta(\Delta)\delta(J_2) + \sum_{\Delta'>0,\ell'} \delta(\Delta -\Delta')\delta(J_2 - J_2') a_{\Delta',\ell'} G_{\Delta',\ell'}(1/2,1/2)
\ea
denote the OPE measure evaluated at $z = \bar{z} =1/2$, and define OPE moments as
\ba
\nu_{m,n} = \int_0^\infty d\Delta \; dJ_2 \;\Delta^m J_2^n \mu(\Delta,J_2).
\ea

In this appendix, we will prove that the double moment sequence $(\nu_{m,n})_{m,n\geq0}$ is Stieltjes determinant, in that it satisfies Carleman's criteria for the double Stieltjes moment problem:

\ba
\sum_{n \geq 1} \nu_{n,0}^{-1/(2n)} = + \infty, \quad\sum_{n \geq 1} \nu_{0,n}^{-1/(2n)} = + \infty.
\label{doublecarlemanscrit}
\ea

\subsection{Moment generating function and Carleman's condition}
\label{mgftocarlemanscrit}
In this subsection, we will prove a lemma which we use in the proof of determinacy for the OPE moment sequence. More specifically, we would like to prove the following: Let $f$ be a positive density on $X = [0,\infty)$ with moments
\ba
m_n = L_f[X^n] =  \int_0^\infty dx \;x^n f(x).
\ea
If the moment generating function $M_X(t) = L_f[e^{Xt}]$ is bounded in some neighborhood $t \in (-t_0,t_0)$ with $t_0>0$, then

\ba
\sum_{n \geq 1} m_{2n}^{-1/(2n)} =  + \infty   \quad \mathrm{and} \quad \sum_{n \geq 1} m_n^{-1/(2n)} =  + \infty
\label{mgftocarlemans}
\ea
hold true. In other words, Carleman's condition for both the Hamburger and Stieltjes problem is implied by the existence of a moment generating function for a measure supported on the positive real line.

\paragraph{Proof} Taylor expand the exponential around $t = 0$ to get
\ba
M_X(t) = L_f[e^{Xt}] = \sum_n \frac{m_n}{n!}t^n.
\ea
If this series has a finite radius of convergence, then, by the Cauchy-Hadamard theorem,
\ba
\limsup_{n \to \infty}\left( \left(\frac{m_n}{n!}\right)^{1/n} \right) = \limsup_{n \to \infty}\left( \left(\frac{m_{\alpha n}}{(\alpha n)!}\right)^{1/(\alpha n)} \right) < \infty
\ea
for all $\alpha\in \mathbb{N}$. After using Sterling's approximation for the factorial, this is equivalent to
\ba
\limsup_{n \to \infty}\left( \frac{m_{\alpha n}^{1/(\alpha n) }}{n}\right)  < \infty.
\ea
This implies $\sup \{ m_{\alpha k}^{1/(\alpha k)}\}_{k\geq n} = O(n)$, or that there exists a constant $c > 0 $ such that $c/n \leq \inf \{ m_{\alpha k}^{-1/(\alpha k)}\}_{k\geq n} \leq m_{\alpha n}^{-1/(\alpha n)} $. Thus,
\ba
c \sum^N_{n = 1} \frac{1}{n} \leq \sum^N_{n=1} m_{\alpha n}^{-1/(\alpha n)} .
\label{carlemansbound}
\ea
Since the LHS series diverges as $N \to \infty$, the RHS also does. Setting $\alpha = 2$ gives Carleman's condition for Hamburger determinacy. To obtain Carleman's condition for Stieltjes determinancy, use the following inequality:
\ba
\left(\sum_{n}^N a_n \right)^r \leq \sum_{n}^N a_n^r 
\ea
for real $a_n >0$ and $0<r<1$ given by eq.~(2.12.2) from~\cite{hardy1988inequalities}. Applying this to the RHS of eq.~(\ref{carlemansbound}) with $r = 1/2$ and $\alpha =1$ gives
\ba
\left(\sum^N_{n=1} m_{ n}^{-1/ n} \right)^{1/2} \leq \sum^N_{n=1} m_{ n}^{-1/(2 n)}.
\ea
Since the LHS diverges as $N\to \infty$, the RHS does as well. This concludes the proof of eq.~(\ref{mgftocarlemans}).

\subsection{OPE moment determinacy}
We will now prove that the double moment sequence $(\nu_{m,n})_{m,n\geq0}$ is determinant. For the sequence of scaling moments, this can be done directly by proving that their moment generating function $M_\Delta(t) = L_\mu[e^{\Delta t}]$ is bounded on some interval $t \in (-t_0,t_0)$. Moreso, in dimensions 1, 2, and 4, we can apply an exponentiated principal series operator to the correlator and evaluate it at $(z,\bar{z}) = (1/2,1/2)$ to produce a moment generating function for $\mu(\Delta)$:
\ba
\left.e^{t \left(\Omega_+ +\frac{d}{2}\right) } \mathcal{G}(z,\bar{z})\right|_{(z,\bar{z}) = (1/2,1/2 )} &=  \int_0^\infty \int_0^\infty d\Delta dJ_2\;\mu(\Delta,J_2) e^{t \Delta }\\
&=M_{\Delta}(t).
\label{mgfcorrespondence}
\ea
In $d=3$ and $d>4$, we can replace the exact operator $\Omega_+$ with the asymptotic one, so we can also establish that this picture is approximately valid for heavy correlators in arbitrary dimensions. 

For this operation to produce a bounded MGF in some neighborhood of $t = 0$, we require the OPE to converge sufficiently quickly. We can split up $M_\Delta(t)$ as
\ba
M_{\Delta}(t) &= \int_0^{\Delta_0} d\Delta \mu(\Delta) e^{\Delta t} + \int_{\Delta_0}^{\infty} d\Delta \mu(\Delta) e^{\Delta t} \\
&= M_{\Delta < \Delta_0}(t) +  M_{\Delta \geq \Delta_0}(t),
\ea
where we have suppressed the integration over $J_2$. Letting $1 \ll \Delta_0 < \infty$, we see $M_{\Delta < \Delta_0}(t) <\infty$ for all $t$ as it is just a finite sum over a smooth function. We now want to show that there exists some $t_0>0$ such that $M_{\Delta \geq \Delta_0}(t)<\infty$ for all $t \in ( -t_0,t_0)$. 

In~\cite{Pappadopulo:2012jk}, the authors showed in the Euclidean section that 
\ba
\sum_{\Delta \geq \Delta_0} a_{\Delta,\ell} G_{\Delta,\ell}(r,\eta) =  O\left(\Delta_0^{2\Df} r^{\Delta_0} \right).
\label{opeconvergence}
\ea
For $\Delta_0 \gg 1$, we can use the heavy block approximation of eq.~(\ref{confblockasym}). Set $\eta =1$ and $r =  r^\star e^{t}$ with $r^\star \equiv (3-2 \sqrt{2}) $. When $t=0$, this is equivalent to evaluating $(z,\bar{z}) = (1/2,1/2)$. We can rewrite the asymptotic conformal block here as
\ba
G_{\Delta,\ell}(r,\eta) \approx f_d(r^\star ,t) G_{\Delta,\ell}(r^\star ,1) e^{\Delta t},
\ea
where $f_d(r,t) = \left(\frac{1-r^2}{1-r^2 e^{2 t}}\right)^{d/2}$ is bounded for $t < -\log \left(r \right)$ and can be brought outside of the OPE sum. Making this replacement in eq.~(\ref{opeconvergence}) and dividing both sides by $f_d(r^\star,t)$, we have
\ba
\sum_{\Delta \geq \Delta_0} a_{\Delta} G_{\Delta}(r^\star,1) e^{\Delta t} =  O\left(\Delta_0^{2\Df} (r^\star e^{t})^{\Delta_0} \right).
\ea

Recognizing the LHS as $M_{\Delta \geq \Delta_0}(t)$, we see that for $t < -\log(r^\star)$, $M_{\Delta \geq \Delta_0}(t) < \infty$ for all $\Delta_0$. Since both $M_{\Delta < \Delta_0}(t),M_{\Delta \geq \Delta_0}(t)<\infty$ for $t < -\log(r^\star)$, $M_{\Delta}(t) < \infty$ for all $t \in (\log(r^\star),-\log(r^\star))$, the moment generating function is well defined. Therefore, by our lemma in (\ref{mgftocarlemanscrit}), the moment sequence $(\nu_{n,0})_{n \geq 0}$ satisfies the condition of eq.~(\ref{doublecarlemanscrit}).

We now want to show that the sequence in spin Casimir moments $(\nu_{0,n})_{n \geq 0}$ satisfies Carleman's condition for the Stieltjes problem:
\ba
\sum_{n\geq1} (\nu_{0,n})^{-\frac{1}{2n}} = + \infty.
\label{sdeternu}
\ea
To do this, we will introduce an auxiliary moment sequence and MGF defined as
\ba
M_{\sqrt{J_2}}(s) = \int_0^\infty \int_0^\infty d\Delta \; dJ_2 \;\mu(\Delta,J_2) e^{s \sqrt{J_2}},
\ea
which generates the moment sequence $\left.\partial_s^{2n}M_{\sqrt{J_2}}(s)\right|_{s=0} = \upsilon_{2n} = \nu_{0,n}$. The idea here is to show that this auxiliary MGF has a finite radius of convergence around $s=0$, so, by the lemma (\ref{mgftocarlemanscrit}), the resulting moment sequence satisfies Carleman's condition for the Hamburger moment problem:
\ba
\sum_{n\geq1} (\upsilon_{2 n} )^{-\frac{1}{2n}} = + \infty,
\ea
which is equivalent to the condition (\ref{sdeternu}) by construction.

The key fact for this proof is the spinning unitarity bound:
\ba
\Delta \geq \ell+d-2 \geq \sqrt{ J_2}.
\label{spinunitaritybound}
\ea
Once again, we split up
\ba
M_{\sqrt{J_2} }(s) &= \int_0^{\Delta_0} \int_0^\infty d\Delta \; dJ_2 \;\mu(\Delta,J_2) e^{s \sqrt{J_2}} + \int_{\Delta_0}^{\infty} \int_0^\infty d\Delta \; dJ_2 \;\mu(\Delta,J_2) e^{s \sqrt{J_2}}\\
&= M^{\Delta < \Delta_0}_{\sqrt{J_2} }(s) +M^{\Delta \geq \Delta_0}_{\sqrt{J_2} }(s).
\ea
By eq.~(\ref{spinunitaritybound}), $M^{\Delta < \Delta_0}_{\sqrt{J_2} }(s)$ is given by a finite sum and is therefore trivially bounded. 

We now just need to show $M^{\Delta \geq \Delta_0}_{\sqrt{J_2} }(s)<\infty$ in some neighborhood of $s = 0$. For $s<0$, we use the fact that $J_2>0$ to bound
\ba
\int_{\Delta_0}^{\infty} \int_0^\infty d\Delta \; dJ_2 \;\mu(\Delta,J_2) e^{s \sqrt{J_2}} \leq \int_{\Delta_0}^{\infty} \int_0^\infty d\Delta \; dJ_2 \;\mu(\Delta,J_2)  < \infty.
\ea
For $s > 0$, we use eq.~(\ref{spinunitaritybound}) to write
\ba
\int_{\Delta_0}^{\infty} \int_0^\infty d\Delta \; dJ_2 \;\mu(\Delta,J_2) e^{s \sqrt{J_2}} \leq \int_{\Delta_0}^{\infty} \int_0^\infty d\Delta \; dJ_2 \;\mu(\Delta,J_2)  e^{s \Delta}  < \infty
\ea
for $s \in (\log(r^\star),-
\log(r^\star ))$. Thus, $M_{\sqrt{J_2} }(s) < \infty$ for $s \in (\log(r^\star),-
\log(r^\star ))$ and 
\ba
\sum_{n\geq1} (\upsilon_{2 n} )^{-\frac{1}{2n}} = \sum_{n\geq1} (\nu_{0,n})^{-\frac{1}{2n}}  = + \infty.
\ea
This concludes the proof.

\section{Bounds on inverse moments}
\label{app:inversebound}

The lower bounds on polynomial moments in $\Delta$ (\ref{momentbound}) allow us to strengthen our bounds on inverse moments $L_{\mu/\{0\}}\left[\frac{1}{\Delta^k}\right]$ for some $k>0$. To do this, we will introduce a regulator $0<\epsilon<\Delta_{\mathrm{gap}}$, and bound $L_{\mu/\{0\}}\left[\frac{1}{\epsilon - \Delta^k}\right]$. First, since $\epsilon < \Delta_{\mathrm{gap}}$, the support of $\mu$ lies above the pole at $\Delta = \epsilon^{1/k}$ and contributions are weighted by the tail of the power law $\sim \frac{1}{\Delta^k}$. Thus,
\ba
\left|  L_{\mu/\{0\}}\left[\frac{1}{\epsilon - \Delta^k}\right] \right|<\infty.
\ea

Now, observe that
\ba
\left | L_{\mu/\{0\}}\left[\frac{1}{\epsilon - \Delta^k}\right] \right| = \left |L_{\mu/\{0\}}\left[\sum_i \epsilon^{-1-i} \Delta^{ik}\right]\right|  =\left |L_{\mu/\{0\}}\left[\sum_i | \epsilon^{-1-i} \Delta^{ik}|\right]\right| <\infty.
\ea
Therefore, we can use Fubini's theorem to swap the power series and measure functional, and bound each term in the sum with the lower bound of eq. (\ref{momentbound}):
\ba
L_{\mu/\{0\}}\left[\frac{1}{\epsilon - \Delta^k}\right]&= L_{\mu/\{0\}}\left[\sum_i \epsilon^{-1-i} \Delta^{ik}\right]\\&=\sum_i \epsilon^{-1-i} L_{\mu/\{0\}}[\Delta^{ik}] \\&> (\nu_0-1) \sum_i \epsilon^{-1-i} \left((\sqrt{2}\Df)^{ik} + O(\Df^{ik-1}) \right)\\&= \frac{\nu_0-1}{\epsilon-(\sqrt{2}\Df)^k} + O(\Df^{-k-1}).
\ea
Taking $\epsilon \to 0$ and multiplying by $-1$ gives
\ba
\frac{L_{\mu/\{0\}}\left[\frac{1}{\Delta^k}\right] }{L_{\mu/\{0\}}[1]}< \frac{1}{(\sqrt{2}\Df)^k}+ O\left(\frac{1}{\Df^{k+1}}\right).
\ea
An equivalent result can be obtained by integrating over the minimal measure $\mu^{(-)}(\Delta)$:
\ba
\frac{L_{\mu/\{0\}}\left[\frac{1}{\Delta^k}\right]}{L_{\mu/\{0\}}\left[1\right]} \lesssim \frac{L_{\mu^{(-)}/\{0\}}\left[\frac{1}{\Delta^k}\right]}{L_{\mu^{(-)}/\{0\}}\left[1\right]} = \frac{1}{(\sqrt{2}\Df)^k} .
\ea

\section{Subleading relations from the diagonal limit}
\label{app:subleading}

Let us consider subleading constraints on the projective scaling dimension moments, which at large $\Delta_{\phi}$ admit an expansion of the form
\ba
\frac{\nu_n}{\nu_0} = \sum_{k = 0}^n a^{(k)}_{n}\Df^{n-k}.
\label{ansatz}
\ea
We can study constraints on the coefficients $a_n^{(k)}$ which arise from the crossing equation by plugging in this ansatz and computing relations order-by-order in $\Df$. The expansion of \ref{ansatz} is truncated at $O(1)$, as any further subleading terms are related to $O(1/\Df)$ corrections from the non-asymptotic piece of the regulated conformal block (see appendix~\ref{app:inversebound}).

To analyze subleading coefficients with $k=1$, we will scale $\Delta \to \xi \Delta$ and $\Df \to \xi \Df$ and study the terms in the constraint coefficient of order $\xi^{\Lambda}$ and $\xi^{\Lambda-1}$. Here, the $k=0$ coefficients that show up in the $\xi^{\Lambda-1}$ term are related to the $k=1$ coefficients in the $\xi^\Lambda$ term. Going to the diagonal $z = \bar{z}$ and taking derivatives with respect to $z$, we find the constraint from crossing at order $\xi^\Lambda$ and $\xi^{\Lambda-1}$ read 
\ba
0&=\xi^\Lambda\left(L_{\mu/\{0\} }\left[\left(2 \sqrt{2} \Delta -4 \Delta _{\phi }\right)^{\Lambda }\right] +(-4 \Delta _{\phi })^{\Lambda } \right) 
\\&+\xi^{\Lambda-1}\left(L_{\mu/\{0\}}\left[\left(2 \sqrt{2} \Delta -4 \Delta _{\phi }\right)^{\Lambda } \left(\frac{\zeta \Lambda  }{4 \left(\sqrt{2} \Delta -2 \Delta _{\phi }\right)}+\frac{(1-\Lambda ) \Lambda  \left(\sqrt{2} \Delta -8 \Delta _{\phi }\right) }{8 \left(\sqrt{2} \Delta -2 \Delta _{\phi }\right)^2} \right)\right] \right.\\
&\left.\qquad\qquad+(1-\Lambda ) \Lambda  \left(-4 \Delta _{\phi }\right)^{\Lambda -1} \right) \\& + O\left(\xi^{\Lambda-2}\right),
\ea
where $L_{\mu/\{0\}}[1] = \nu_0 - 1$ and $\zeta \equiv \left(3-2 \sqrt{2}\right) d$. While we have written these constraints separately, they are indeed related at subleading order $\Df^{\Lambda-1}$ after plugging in our ansatz (\ref{ansatz}) and setting $\xi=1$.

Expanding each of the terms as a polynomial and replacing $L_{\mu/\{0\}}[\Delta^\lambda] = \nu_\lambda$ for $\lambda>0$ and $L_{\mu/\{0\}}[1] = \nu_0-1$ , we find the constraint 
\ba
0= \sum^\Lambda_{\lambda = 0} &\left[\binom{\Lambda}{\lambda } \left(2 \sqrt{2}\right)^{\lambda }  \frac{\nu_\lambda}{\nu_0} \left(-4 \Delta _{\phi }\right)^{\Lambda-\lambda }  \right.
\\&\,+ 2^{\frac{3 \Lambda }{2}-3} \left(-\sqrt{2} \Delta _{\phi }\right)^{-\lambda +\Lambda -1} \left( \binom{\Lambda }{\lambda }\frac{\nu _{\lambda -1}}{\nu_0}(\lambda -1) \lambda  \Delta _{\phi } \right.\\
&\left.\left.\qquad\qquad\qquad\qquad\qquad\qquad+\binom{\Lambda}{\lambda,\Lambda-\lambda-1}\frac{\nu _{\lambda }}{\nu_0} \sqrt{2}   (\zeta +2 \lambda -2 \Lambda +2)\right)\right] \\ +\,\zeta & 2^{2\Lambda -3}  \Lambda \frac{\Df^{\Lambda-1}}{\nu_0} + O\left(\Df^{\Lambda - 2}\right),
\ea
where the first term in the last line is a non-projective constraint, involving a factor not of the form $\nu_\lambda/\nu_0$. The remaining terms are all projective and can be expanded with the ansatz (\ref{ansatz}). Doing so, we recover relations between the $a_n^{(k)}$ coefficients at order $\Df^{\Lambda}$ and $\Df^{\Lambda-1}$:
\ba
0 =\,& \Delta _{\phi }^{\Lambda } \sum^\Lambda_{\lambda =0} \binom{\Lambda }{\lambda }(-1)^{\Lambda -\lambda } 2^{2 \Lambda -\frac{\lambda }{2}} a_\lambda^{(0)} \\ & + \Df^{\Lambda-1}\left( \sum^\Lambda_{\lambda=0}  \left[\binom{\Lambda}{\lambda}(-1)^{\Lambda -\lambda } 2^{-\frac{\lambda }{2}+2 \Lambda -4} \mathfrak{A}(\Lambda,\lambda)\right] + \zeta 2^{2\Lambda-3} \frac{\Lambda}{\nu_0} \right) \\&+ O\left( \Df^{\Lambda-2} \right)
\ea
with 
\ba
 \mathfrak{A}(\Lambda,\lambda) =\left(-2 (\lambda -\Lambda ) a_\lambda^{(0)} (\zeta +2 \lambda -2 \Lambda +2)-16 a_\lambda^{(1)}+\sqrt{2} (\lambda -1) \lambda  a_{\lambda-1}^{(0)}\right).
\ea
Note that since $\nu_0/\nu_0 = 1$ with no subleading terms, we have $a_0^{(0)} = 1$ and $a_0^{(k>0)}=0$. 

To check this relation, let $\Lambda = 1$. At order $\Df$, we have $a_1^{(0)} = \sqrt{2}$, and at order 1 we have $a_1^{(1)} = -\frac{\zeta  \left(\nu _0-1\right)}{4 \sqrt{2} \nu _0} = \frac{1}{8} \left(4-3 \sqrt{2}\right) d \frac{  \left(\nu _0-1\right)}{ \nu _0}$. This agrees with our previous bound (\ref{firstbound}) obtained without these general relations. 

\end{appendix}

\bibliography{biblio}{}
\bibliographystyle{JHEP}
\end{document}